\documentclass[structabstract]{aa}
\usepackage{txfonts}
\usepackage{graphicx}
\usepackage{lscape}

\begin{document}

\title{Compact sources in the Bologna Complete Sample: \\
       high resolution VLA observations and optical data.}

\subtitle{}

\author{E. Liuzzo\inst{1}, S. Buttiglione\inst{2}, G. Giovannini\inst{1,3}, M. Giroletti\inst{1}, S. Capetti\inst{4}, G. B. Taylor\inst{5,6}}

\offprints{E.Liuzzo, \email{liuzzo@ira.inaf.it}}

\institute{INAF - Istituto di Radioastronomia, via Gobetti 101, 40129 Bologna, Italy.
\and INAF - Osservatorio Astronomico di Padova, vicolo dell’Osservatorio 5, 35122 Padova, Italy
\and Dipartimento di Astronomia, Universit\'a di Bologna, via Ranzani 1 , 40127 Bologna, Italy
\and INAF - Osservatorio Astronomico di Torino, via Osservatorio 20, 10025 Pino Torinese, Italy
\and Department of Physics and Astronomy, University of New Mexico, Albuquerque NM 87131, USA 
\and also Adjunct Astronomer at the National Radio Astronomy Observatory, USA}

\date{Received ........ / Accepted .........}

\abstract {Among radio galaxies, compact sources are a 
class of objects not yet well understood, and most of them cannot be included
in classical populations of compact radio sources (flat spectrum AGN or compact steep spectrum sources).}
   {Our main goal is to analyze the radio and optical properties of 
a sample of compact sources and compare them with FRI/FRII extended radio galaxies.} 
   { We selected in the Bologna Complete Sample a sub sample of 
Compact sources, naming it the C BCS sample. We collected new and literature 
sub-arcsecond resolution multi-frequency VLA images and optical data. 
We compared total and nuclear radio power with optical emission line 
measurements.}
   {The [OIII] 
luminosity $-$ 408 MHz total power relation found in High and Low excitation
galaxies, as well as in young (CSS) sources, holds also for the C BCSs. However, C BCSs present
 higher [OIII] luminosity than
expected at a given total radio power, and they show the same 
correlation of Core Radio Galaxies, but with a higher radio power. }
   {C BCSs appear to be the high power tail of Core Radio 
Galaxies.
For most of the C BCSs, the morphology seems to be strongly dependent to the 
presence of dense environments (e.g. cluster or HI-rich galaxies) and to a low
age or restarted radio activity.}


\maketitle

\section{INTRODUCTION.} \label{Sec_Intro}
According to their radio power and morphologies, radio galaxies are classified
as FRI and FRII (Fanaroff $\&$ Riley 1974) and Compact sources. FR I and FR II
are extended sources on kpc up to Mpc scale. Their properties have been 
analyzed by several authors (e.g. Fanaroff $\&$ Riley 1974, Laing et al. 1983, Fanti et al. 1986,
Ledlow $\&$ Owen 1996). Recently, Capetti et al. 2011 (and 
references therein) discussed the correlation between optical and radio 
properties at the light of unification models and accretion properties.

Sources with a projected linear size smaller than 15-20 kiloparsec are 
usually defined as compact sources and can be 
high or low power radio sources. 
High radio power sources can have flat or steep spectrum. Flat spectrum 
sources are small because of 
projection and relativistic effects being identified (in agreement with 
unified models, see e.g. Urry $\&$ Padovani 1995) as objects dominated by the 
emission of a relativistic jet oriented at a small angle with respect to the 
line of sight. According to their radio power and optical properties, they 
are classified as FSRQ (Flat Spectrum Radio Quasars) or BL Lac objects.

High power, compact steep spectrum (CSS) sources can be small because they are 
young sources (e.g. Stanghellini et al. 2005 and references therein).
CSS sources are not beamed sources and they are likely to be young radio 
galaxies that could evolve into large radio objects, FR I/FR II (see Fanti et 
al. 1995; Readhead et al. 1996; O'Dea 1998, for reviews, but also van Breugel 
et al. 1984). Strong support to this scenario comes from the measurements of 
proper motions 
of the hot spots of some of them (Polatidis $\&$ Conway 2003, Giroletti $\&$ 
Panessa 2009), with separation velocities of 0.1-0.4 h$^{-1}$ c, and then 
small kinematic ages which are in agreement with spectral ages derived from 
flux density measurements (Murgia et al. 1999; Murgia 2003). 
Kunert-Bajraszewska et al. 2010 selected and studied the 
properties of low power CSS sources
with flux density $<$ 70 mJy at 1.4 GHz and $\alpha^{4.85GHz}_{1.4GHz}$ 
$>$ 0.7. These authors suggest the 
existence of a large population of short-lived objects, poorly known. Some of
these could be precursor of large-scale FR I galaxies.

To better investigate the nature and properties of compact radio
galaxies, we selected from the Bologna Complete Sample (BCS, Giovannini et
al. 2001, 2005, Liuzzo et al. 2009b) all sources with a projected linear size smaller than 20 
kpc. We will name this
sub-sample the C BCS (Compact BCS sources) and it is composed by 18 objects. 
This complete sub-sample is selected at low frequency, therefore should present
no bias with respect to the source orientation, possible beaming effects and
spectral index.
We note that most sources in our sample show a moderately steep spectral 
index, and could not be included in samples as the one presented by 
 Kunert-Bajraszewska et al. 2010. 

The radio power of C BCS sources is low with respect to powerful FR II or 
FSRQ, but in the same range of giant FR I radio galaxies (10$^{23-26}$W/Hz at 
408 MHz).  
Some of them were previously analyzed by us in the radio band (Giroletti et 
al. 2005b). We present in this paper radio data of the 5 remaining ones with 
new high resolution Very Large Array (VLA) observations. We also discuss 
their optical properties, showing for the first time their optical spectra, 
since the emission lines analysis is fundamental in order to understand the 
nature and properties of these objects.

Moreover we include in this paper data of few extended BCS 
sources in order to usefully compare the C BCS radio and optical properties  with 
different radiogalaxy types (e.g. FRII and FRI extended sources).

The layout of the paper is the following: \\
- in Sect. 2, we describe radio and optical data for C BCS sample, in particular the new 
high resolution VLA images and TNG (Telescopio Nazionale Galileo) observations;\\
- in Sect. 3, we present optical results for all our targets;\\
- in Sect. 4, we analyze the relation between the optical and radio emissions;\\
- in Sect. 5, we report notes on single sources ;\\
- in Sect. 6, we discuss our main results with literature ones;\\
- in Sect. 7, we resume our main conclusions.

Throughout this paper, we use of H$_{0}$ = 70 km s$^{-1}$ Mpc$^{-1}$, 
$\Omega_{\rm M} = 0.3$ and $\Omega_\Lambda = 0.7$. 
Spectral indices are defined such that $S(\nu) \propto \nu^{-\alpha}$.

\section{RADIO AND OPTICAL DATA.} \label{Sect_Radio}

Table 1 summarizes references for radio and optical data of 
all C BCS;
N indicates sources for which new data are presented for the first time in 
this paper. 
We added also a few extended BCSs with new data 
presented here, to increase the statistics
in the comparison between compact and extended sources.

\subsection{New Radio Data}

In this Section, we describe the data analysis done for the new sub-kpc scale VLA data 
obtained by us for B2 0149+35, B2 0708+32B, B2 0722+30, B2 1254+27, and 
B2 1557+26. We report also the new sub-kpc VLA observations of two peculiar 
BCS radio sources, B2 0331+39 and B2 1512+30. 

The new high resolution VLA observations were obtained in two observing runs 
in 2006 March 11 and 2006 April 04. The array was in A configuration, and the 
observing frequencies were 8.4 GHz and 22 GHz. 
Standard observing schedules for high frequency observations were prepared, 
including scans to 
determine the primary reference pointing, and using a short (3 s) integration 
time and fast switching mode (180 s on source, 60 s on calibrator) for K band 
(22 GHz) scans. 
Post-correlation processing and imaging were performed with the NRAO (National Radio Astronomy Observatory)
Astronomical Image Processing System (AIPS). Parameters of natural 
uv-weighted images are reported in Table 2.   
Our new VLA images for resolved sources are shown in Sect.5.
To separate the different source components and relative fluxes, dimensions, we used in AIPS the JMFIT task which is a sophisticated least-squares fit of image with Gaussian components. Taking into account also spectral index considerations, we identified the core as the unresolved (point-like, see Tab.5) component having the higher flux density in the source.
If given, the spectral index maps are made with AIPS using the same uv-range. 
Typical errors for flux density measurements are $\sim$3$\%$ for 8.4 GHz, and 
$\sim$10$\%$ for 22 GHz.

\begin{table}[th!]
\begin{center}
\caption {{\it Upper:} {\bf C BCS sources} and
{\it Down:} {\bf Extended  (E BCS) sources} with new data published in this 
paper} \label{tabossTOT}
\begin{tabular}{cccc}
\hline
\hline
Name    &Type& Radio &Optical \\
        &    & data  & data \\
\hline
\hline
B2 0116+31 &C BCS& 3 &  N \\
B2 0149+35 &C BCS& N &  10\\
B2 0222+36 &C BCS& 5 &  N    \\
B2 0258+35 &C BCS& 5 &  N      \\
B2 0648+27 &C BCS& 5 &  N    \\
B2 0708+32B&C BCS& N &  -    \\
B2 0722+30 &C BCS& N &  11      \\
B2 1037+30 &C BCS& 5 &  N      \\
B2 1101+38 &C BCS& 6 &  12     \\
B2 1217+29 &C BCS& 4 &  13   \\
3C 272.1   &C BCS& 1 &  14        \\
B2 1254+27 &C BCS& N &  15        \\
B2 1257+28 &C BCS& 9 &  15     \\
B2 1322+36B&C BCS& 1 &  15         \\
B2 1346+26 &C BCS& 7 &  16    \\
3C 305     &C BCS& 8 &  15        \\
B2 1557+26 &C BCS& N &  15   \\
B2 1855+37 &C BCS& 5 &  N     \\
\hline
\hline                                                         
B2 0331+39  & E BCS&N& -\\ 
B2 0844+31  & E BCS&2 &N\\		
B2 1003+35  & E BCS&2&N\\				
B2 1144+35  & E BCS&1&15\\		
B2 1512+30  & E BCS&N&N\\ 
B2 1626+39  & E BCS&1& 14\\
\hline
\hline
\end{tabular}
\end{center}
{\scriptsize N: New high resolution VLA and/or optical data; 1: Giovannini et 
al. 2001; 2: Giovannini et al. 2005: 3: Giroletti et al. 2003; 4: Giroletti et 
al. 2005a; 5: Giroletti et al. 2005b; 6: Giroletti et al. 2006; 7:Liuzzo et 
al. 2009a; 8: Massaro et al. 2009; 9.Liuzzo et al. 2010 ; 10:Crawford et al. 
1999 , 11: Morganti et al. 1992, 12: Capetti et al. 2010, 13: Ho et al. 1997, 
14: Buttiglione et al. 2009, 15: SDSS DR7, Abazajian et al. 2009, 16: Anton 
1993.}
\end{table}

\begin{table*}[th!]
\caption{ {\bf Radio observation parameters} for BCS sources: compact (upper) 
and extended (low)}\label{tabIm}
\begin{center}
\footnotesize
\begin{tabular}{cccccccc}
\hline
\hline
&&8.4 GHz&&&22 GHz&&\\
Name&beam&noise&peak&beam&noise&peak\\
&($^{\prime \prime}\times^{\prime \prime},^{\circ}$)&(mJy/beam)&(mJy/beam)&($^{\prime \prime} \times ^{\prime \prime},^{\circ}$)&(mJy/beam)&(mJy/beam)\\
\hline 
\hline
B2 0149+35   &0.35$\times$0.30, -86 &0.05 &  5.4  &0.12$\times$0.10, +85 &0.09   &  4.2\\
B2 0708+32B  &0.36$\times$0.30, +60 &0.03 & 12.3  &0.12$\times$0.11, +46 &0.06   &  7.3\\
B2 0722+30   &0.34$\times$0.32, -88 &0.02 & 14.6  &0.09$\times$0.08, +71 &0.08   &  5.1\\
B2 1254+27   &0.31$\times$0.30, -56 &0.02 &  1.0  &0.12$\times$0.12, +0  &0.06    &  5.2\\
B2 1557+26   &0.32$\times$0.32, +60 &0.03 & 13.1  &0.12$\times$0.11, +34 &0.06   & 12.5\\
\hline
\hline
B2 0331+39   &0.34$\times$0.29, -43 &0.08 &146.7  &0.13$\times$0.11, -52 &0.07   &117.1\\
B2 1512+30   &0.20$\times$0.20, +40 &0.05 & ND    &0.07$\times$0.08, +23 &0.06   &  ND \\
\hline
\hline
\end{tabular}
\end{center}
{\scriptsize ND indicates sources non detected in our new VLA data.}
\end{table*}

\subsection{Optical Data} \label{Sect_Optical}

We collected optical information available for all C BCS sources,  
presenting new optical spectra and completing with the emission line 
measurements from literature. In Tab.1 , we list references used for our optical analysis. With numbers, we indicate 
literature references for optical information. With N we indicate sources for which optical observations were taken by us 
with the TNG, a 3.58 m optical/infrared 
telescope located on the Roque
de los Muchachos in La Palma Canary Island (Spain). Spectra for five C BCS objects 
are available from the Sloan Digital Sky Survey (SDSS) database, Data Release 
7 (Abazajian et al. 2009). We add to the sample unpublished
optical observations for some extended BCSs to compare the compact BCS 
sources with the extended ones. It is 
important to note that we are considering all optical information available 
up to now for the whole BCS sample.

\subsubsection{TNG observations.}

Tab. 3 contains the journal of observations and basic information for our
targets.  The TNG observations were made using the DOLORES (Device
Optimized for the LOw RESolution) or LRS (Low Resolution Spectrograph) spectrograph installed at the Nasmyth B
focus of the telescope.  The detector is a Loral thinned and back-illuminated
2048x2048 CCD. Observations were performed in service mode during several
nights between August 21 and December 18 2003. The seeing, measured from
  the acquisition images, varied between 0.8 arcsec and 1.3 arcsec. We used the
low resolution LR-B grism and the 1.0 arcsec slit width.  The dispersion is
2.8 \AA\ per pixel and the spectral resolution about 11 \AA.  The typical
useful spectral range is $\sim3500$ to $\sim8000$ \AA\ with increasing
fringing beyond 7200 \AA.  This range of wavelengths enables us to analyze all
the relevant emission lines of the optical spectrum: H$\beta$,
[OIII]$\lambda$$\lambda$ 4959,5007 \AA, [OI]$\lambda$$\lambda$ 6300,64 \AA,
H$\alpha$, [NII]$\lambda$$\lambda$ 6548,84 \AA, [SII]$\lambda$$\lambda$
6716,31 \AA.

\subsubsection{Data analysis}
\label{Sect_odata}

The data were reduced using the LONGSLIT package of NOAO's (National Optical Astronomy Observatory) IRAF\footnote{IRAF 
(Image Reduction and Analysis Facility)} is distributed by the National 
Optical Astronomy Observatories, which are operated by the Association of 
Universities for Research in Astronomy, Inc., under cooperative agreement 
with the National Science Foundation. It is available at 
http://iraf.noao.edu/ reduction software. 
A bias frame was subtracted from any frame, then the flat field correction 
was applied to remove the pixel-to-pixel gain variations. After that, the 
wavelength calibration, the optical distortions corrections and the background
subtraction were applied. One-dimensional spectra were extracted by summing 
in the spatial direction
over an aperture corresponding to the nuclear part of the source: we 
extracted and summed the 6 pixel rows closest to the center of the spectrum, 
corresponding to 1.65\arcsec.
Lastly the relative flux calibration was made using spectro-photometric
standard stars observed during each night.
In Fig. 1 we present the C BCS sources optical spectra after 
the calibration.

In order to properly measure the emission lines intensities, we needed to
subtract the stellar emission of the host galaxies. Before removing the
  stellar continuum, we corrected for reddening due to the Galaxy
  (Burstein et al. 1982, 1984) using the extinction law of
  Cardelli et al. 1989. The galactic extinction E(B-V) used for each object was
  taken from the NASA Extragalactic Database (NED) database. The adopted
method consists on modeling the nuclear spectra with a single stellar
population taken from the Bruzual et al. 2003 library and then subtracting the
best fit model spectrum from the nuclear one.  The templates assume a Salpeter 
Initial Mass Function (IMF) formed in an instantaneous burst, with
solar-metallicity stars in the mass range $0.1 \leq M \leq 125$
M$_{\odot}$. The parameters free to vary independently each other in order to
obtain the best fit are the stellar age (from 1 to 13 Gyr), the metallicity
(from 0.0008 to 0.5 solar metallicity), the normalization of the model, the
velocity dispersion, the continuum emission from the AGN (Active Galactic Nucleus)  and its slope.  Even
if this method of stellar removal gives also an estimate of the velocity
dispersion, for the resolution of our spectra this value is dominated by the
instrumental broadening.  The spectral regions chosen for the fit are centered
on the H$\beta$ and H$\alpha$ emission lines, with a range of 3600 - 5500 \AA\
for the H$\beta$ and 5700 - 7100 \AA\ for the H$\alpha$. The emission lines
are excluded from the fit, since they are strongly affected by the nuclear
emission more than the stellar one.  Other small regions are excluded because
of telluric absorption, cosmic rays or other kind of impurities.  At the end
of this operation, as a result, we obtained the non stellar nuclear emission
produced by the AGN activity.  In Fig. 2 we show as an example of the adopted
procedure, the spectra of 2 C BCS sources. The source spectra are in solid
lines, the top spectra are before the stellar removal and the bottom spectra
are the results of the host galaxy stellar population subtraction. The dotted
line through the top spectra indicates the single stellar population model.
The dashed line across the bottom spectra indicates the zero flux level.
These spectra have a quite flat continuum emission with the overlap of
emission lines produced by the photoionised gas.

\begin{table}[th!]
\begin{center}
\caption{{\bf Log of the new TNG observations}} \label{tab_logOpt}
\label{tabossTNG}
\begin{tabular}{l c c  c c}
\hline\hline
Name & redshift & Date & T$_{exp}$ \\
\hline
\hline	    	 	                  
B2 0116+31    & 0.059    & 2003/08/22 & 1800x2  \\
B2 0222+36    & 0.033    & 2003/08/22 & 2400      \\
B2 0258+35A  & 0.017   & 2003/08/22 & 2400     \\	
B2 0648+27    & 0.041    & 2003/09/22 & 2400     \\
B2 1037+30    & 0.091    & 2003/12/17 & 1800x2 \\
B2 1855+37    & 0.056    & 2003/08/21 & 1800x2 \\
\hline
\hline
B2 0844+31B  & 0.067      & 2003/11/14  & 2400     \\
B2 1003+35    & 0.099       & 2003/11/18  & 1800x2 \\
B2 1512+30    & 0.094        & 2003/08/23  & 1800x2 \\
\hline	     
\hline	    	
\end{tabular}
\end{center}
{\scriptsize In Col.1 there are target names.
Upper sources are C BCS while sources below the two solid lines are extended 
BCSs.
In Col.2 there is the redshift of the sources.
In Col. 3 we report the date of observation and in 
Col. 4 the exposure time in sec where x2 means that the image was repeated.}
\end{table}

\begin{figure*}[th!] 
\centering
\includegraphics[width=0.35\textwidth]{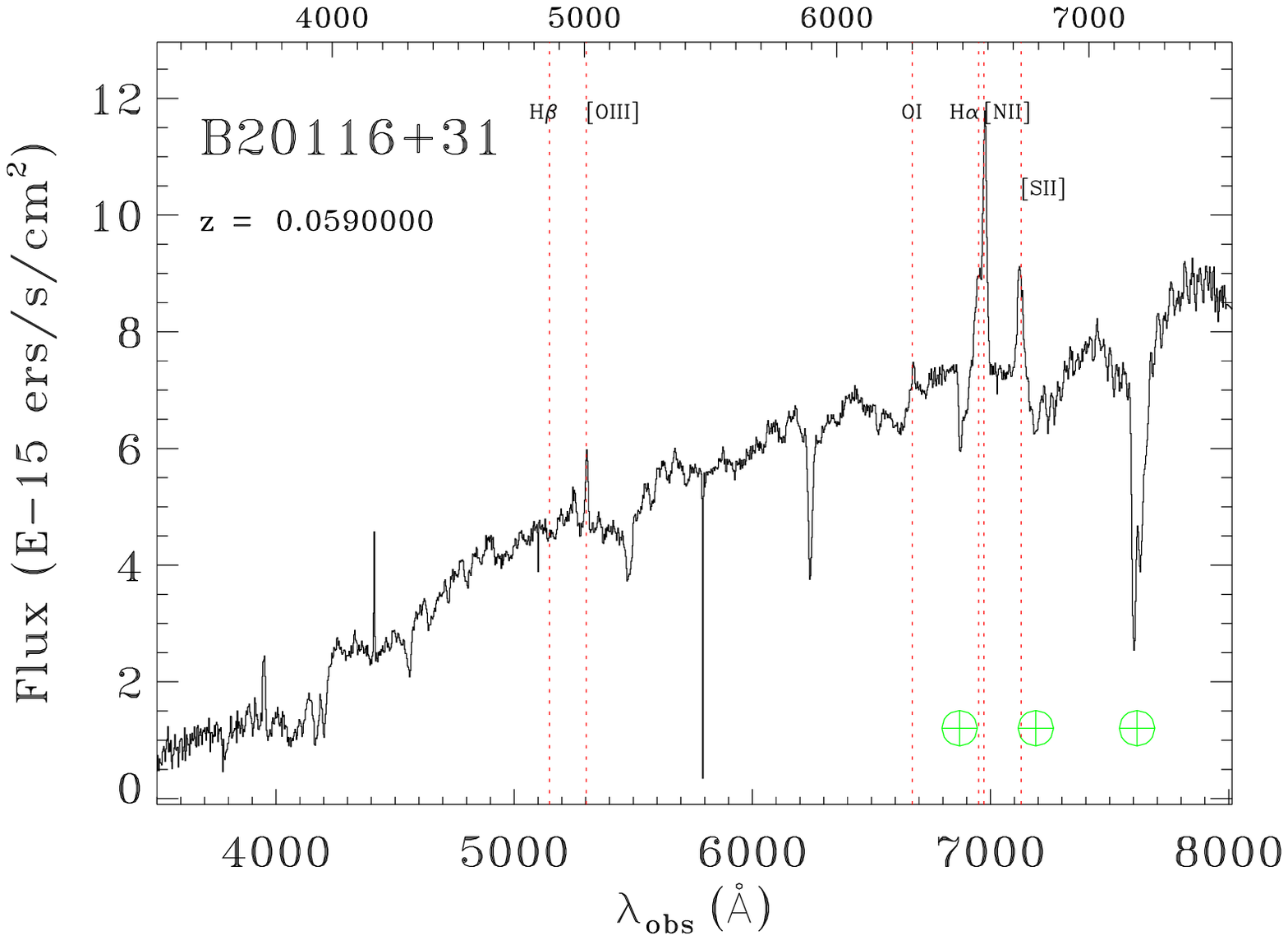}
\includegraphics[width=0.35\textwidth]{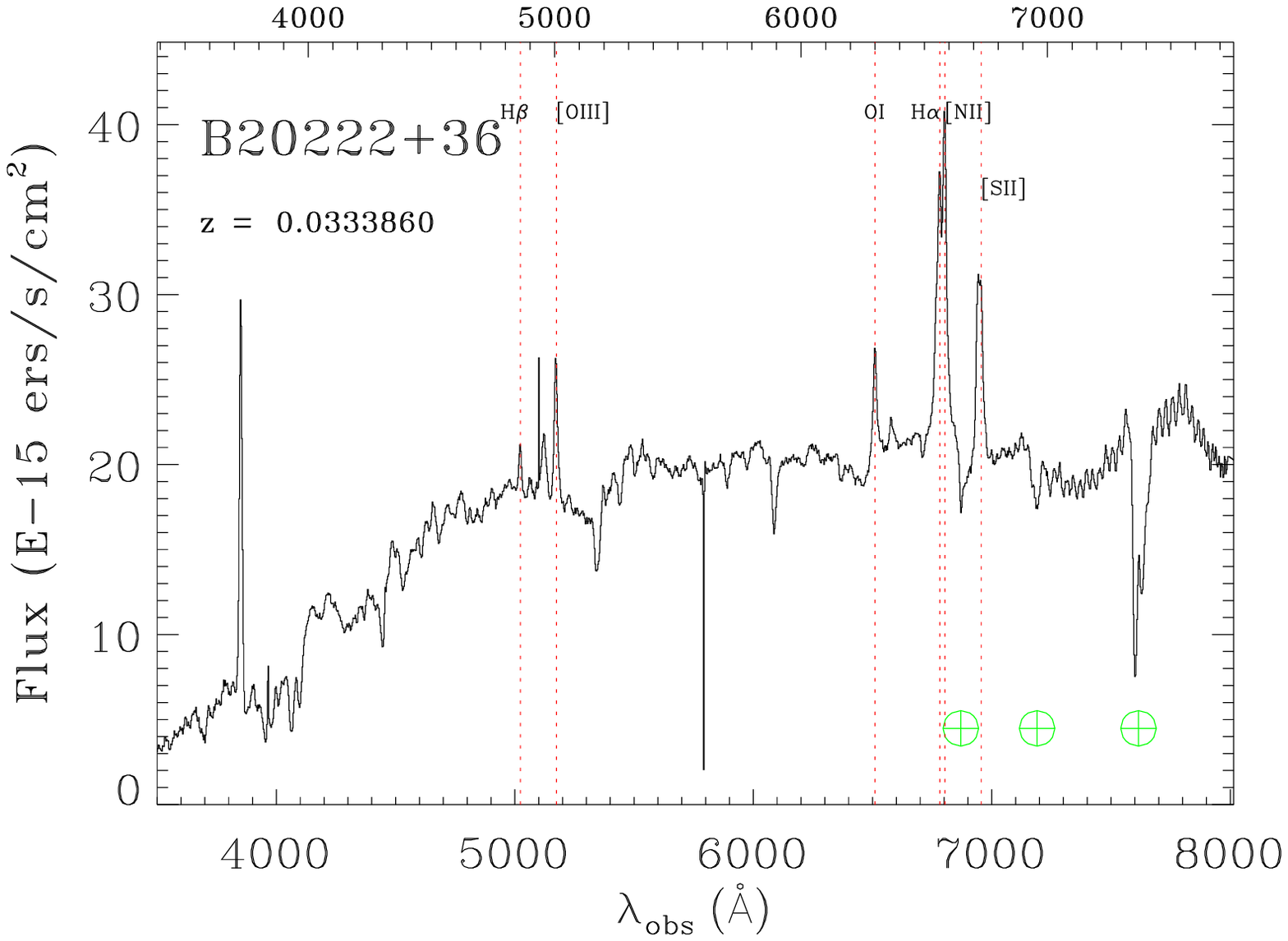}
\includegraphics[width=0.35\textwidth]{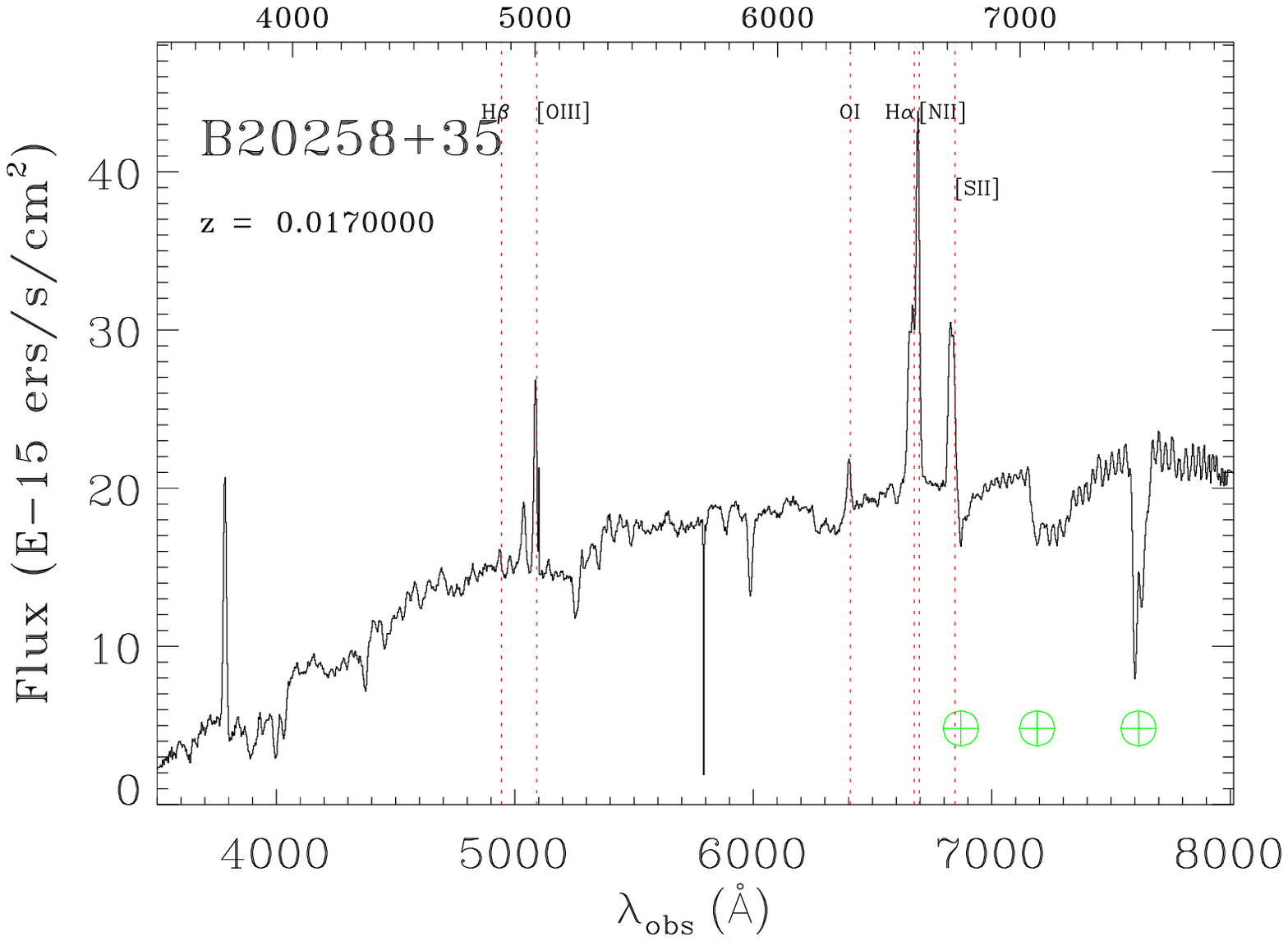}
\includegraphics[width=0.35\textwidth]{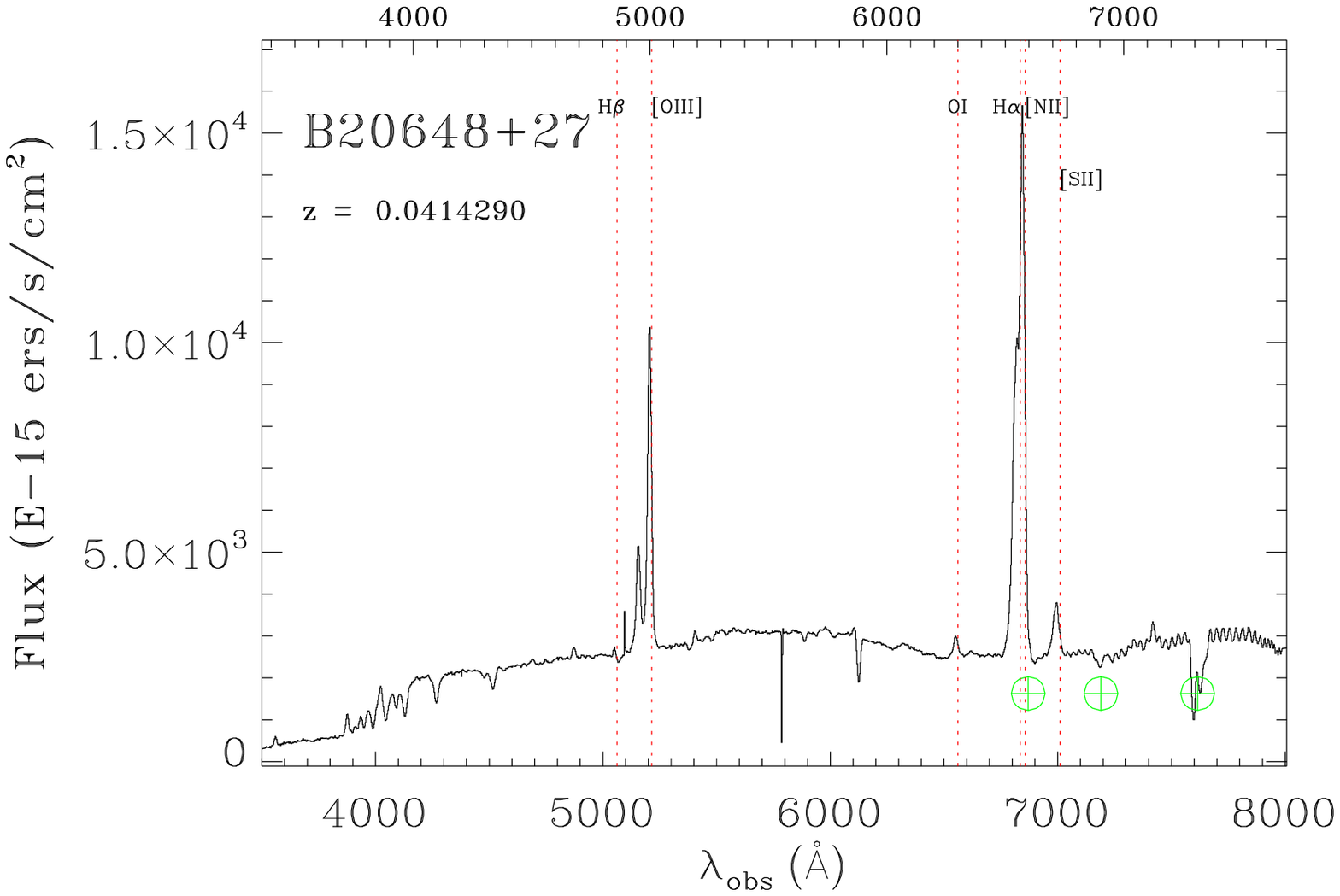}
\includegraphics[width=0.35\textwidth]{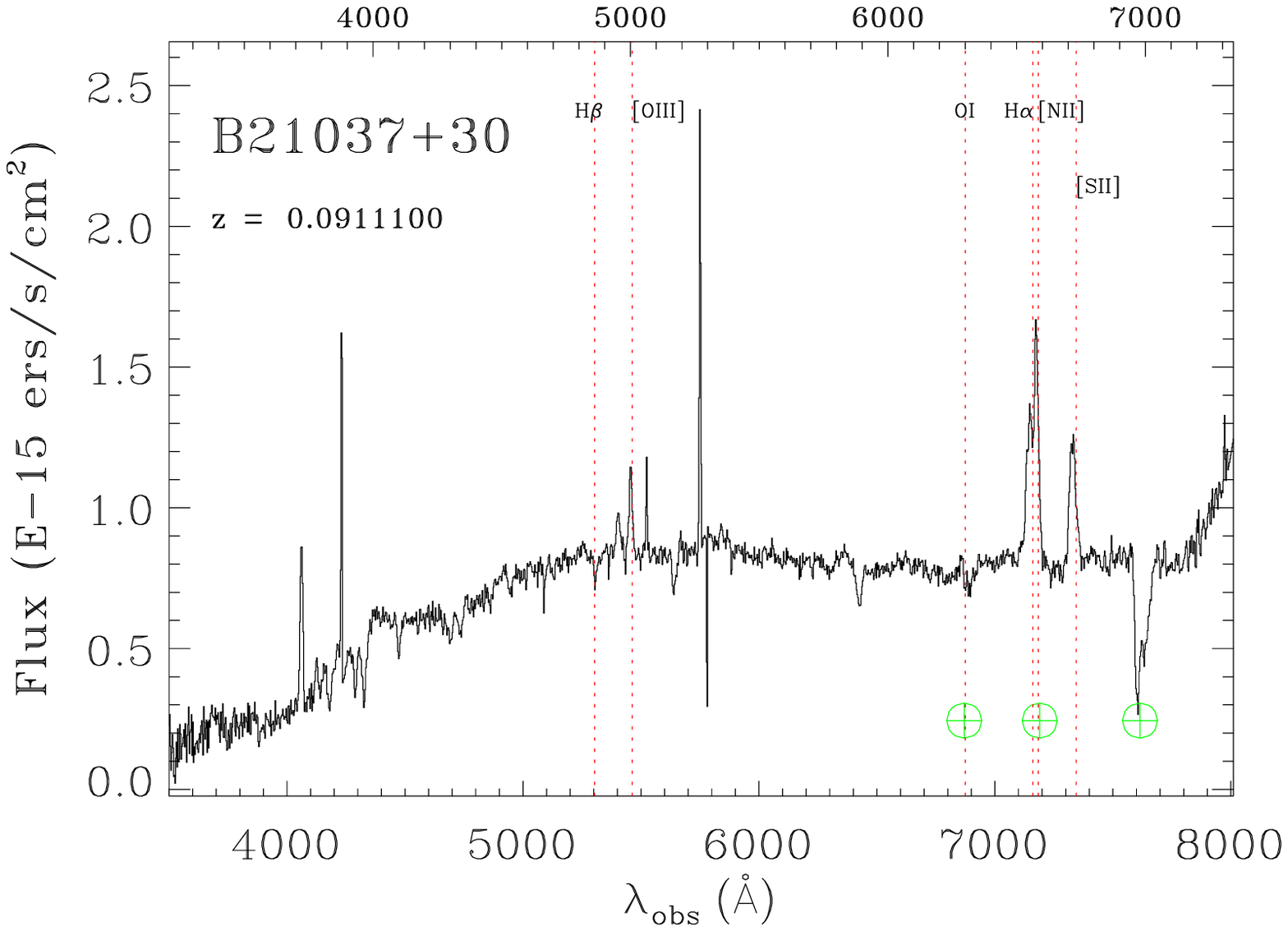}
\includegraphics[width=0.35\textwidth]{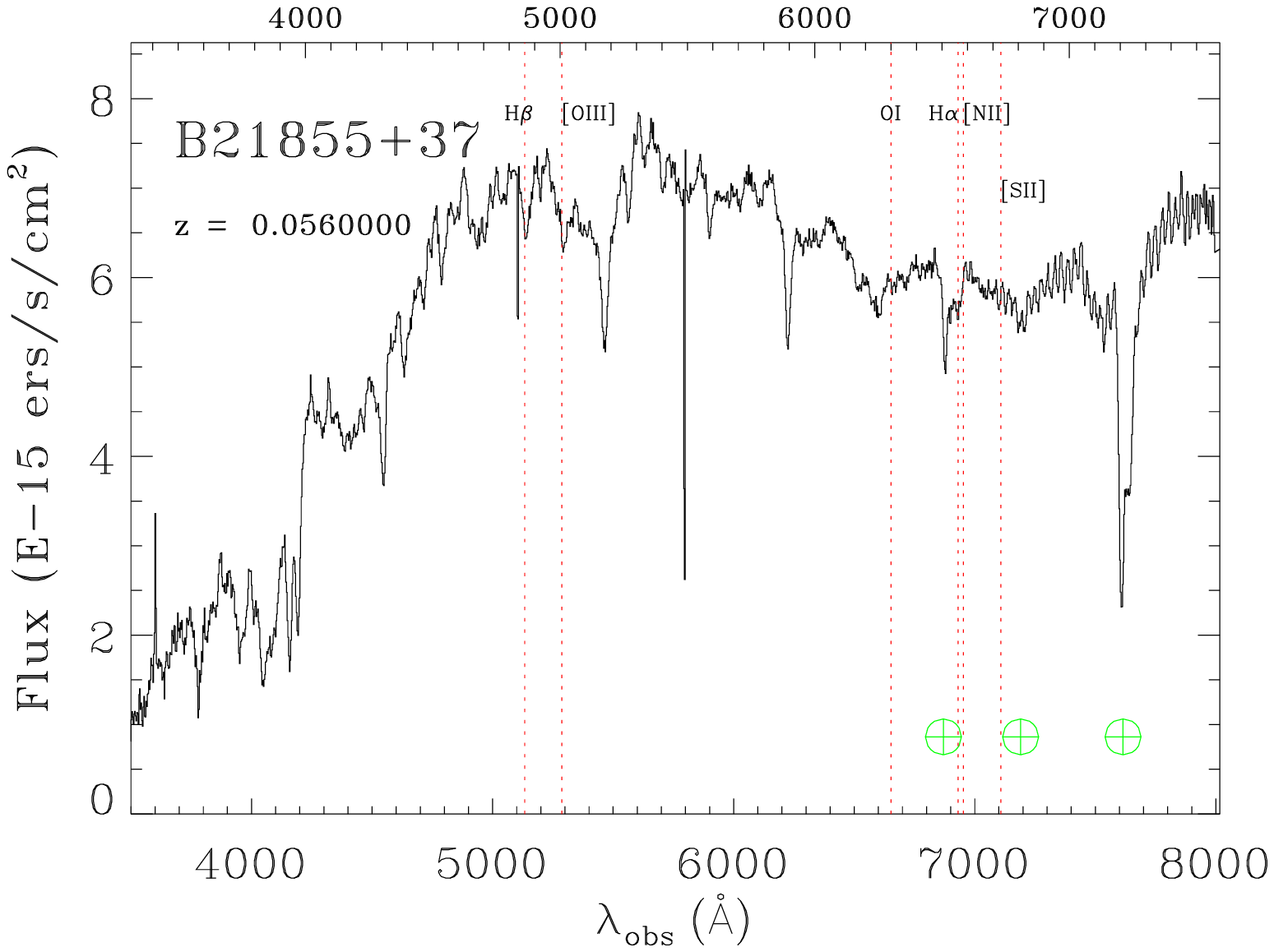}
\includegraphics[width=0.35\textwidth]{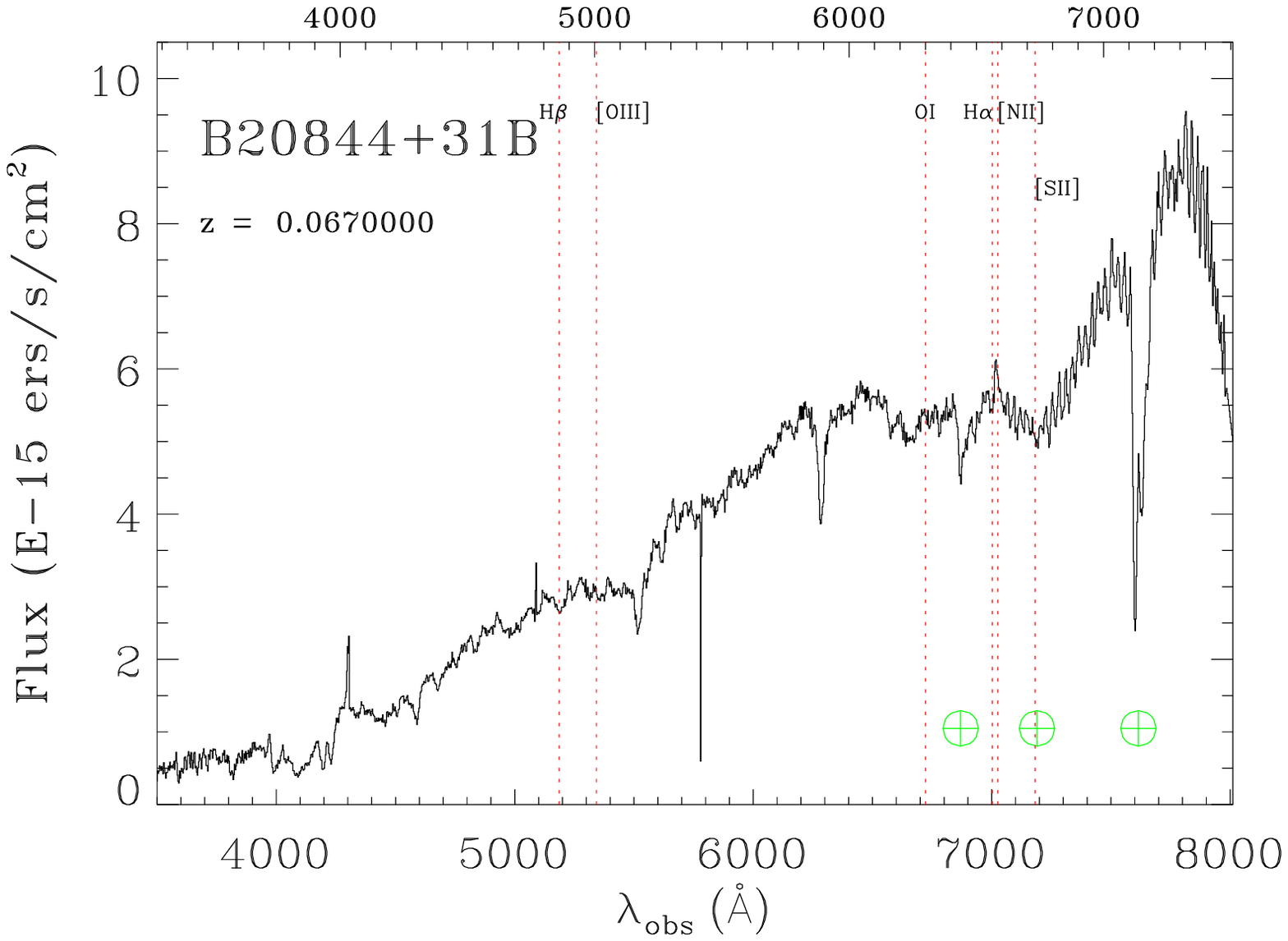}
\includegraphics[width=0.35\textwidth]{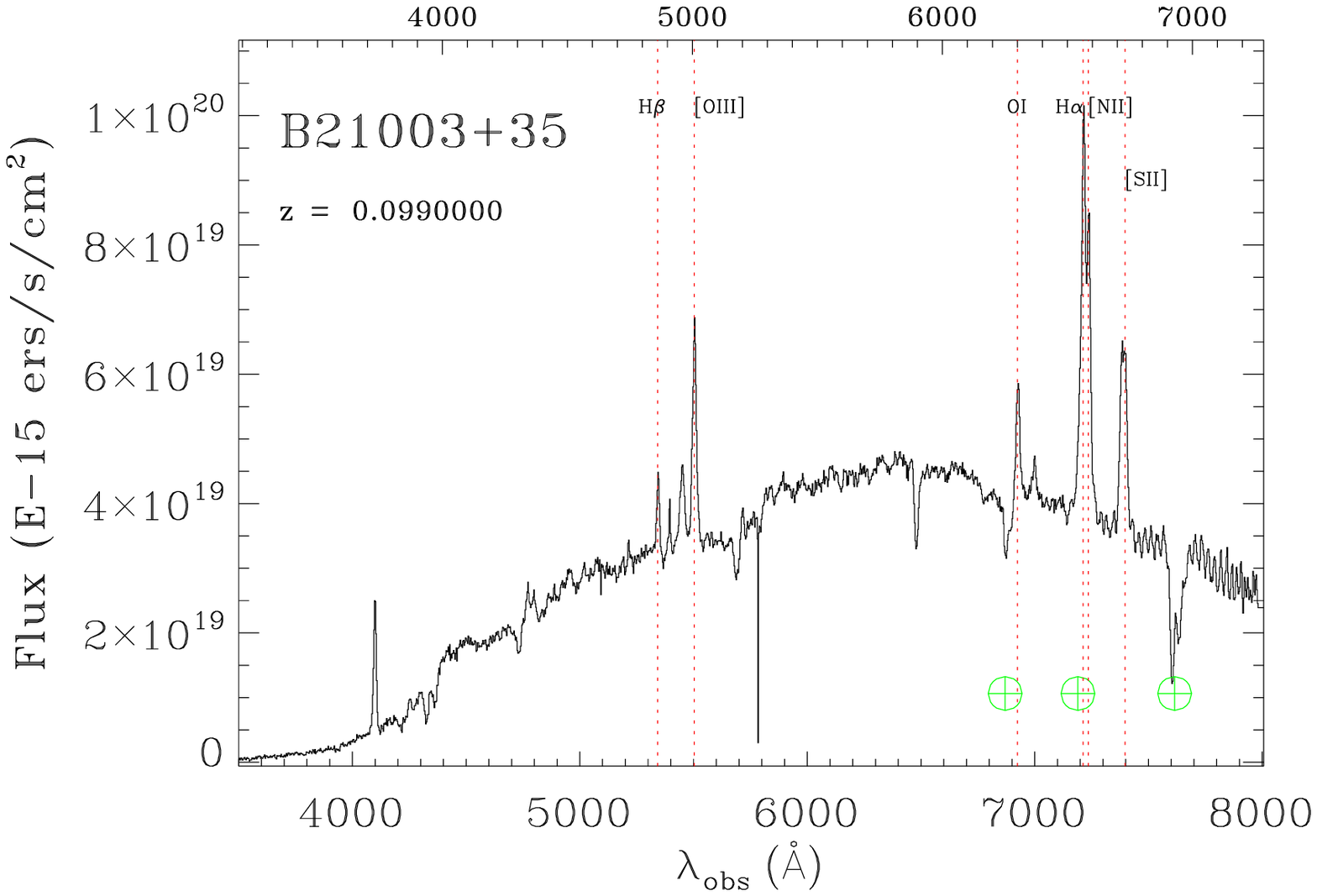}
\includegraphics[width=0.35\textwidth]{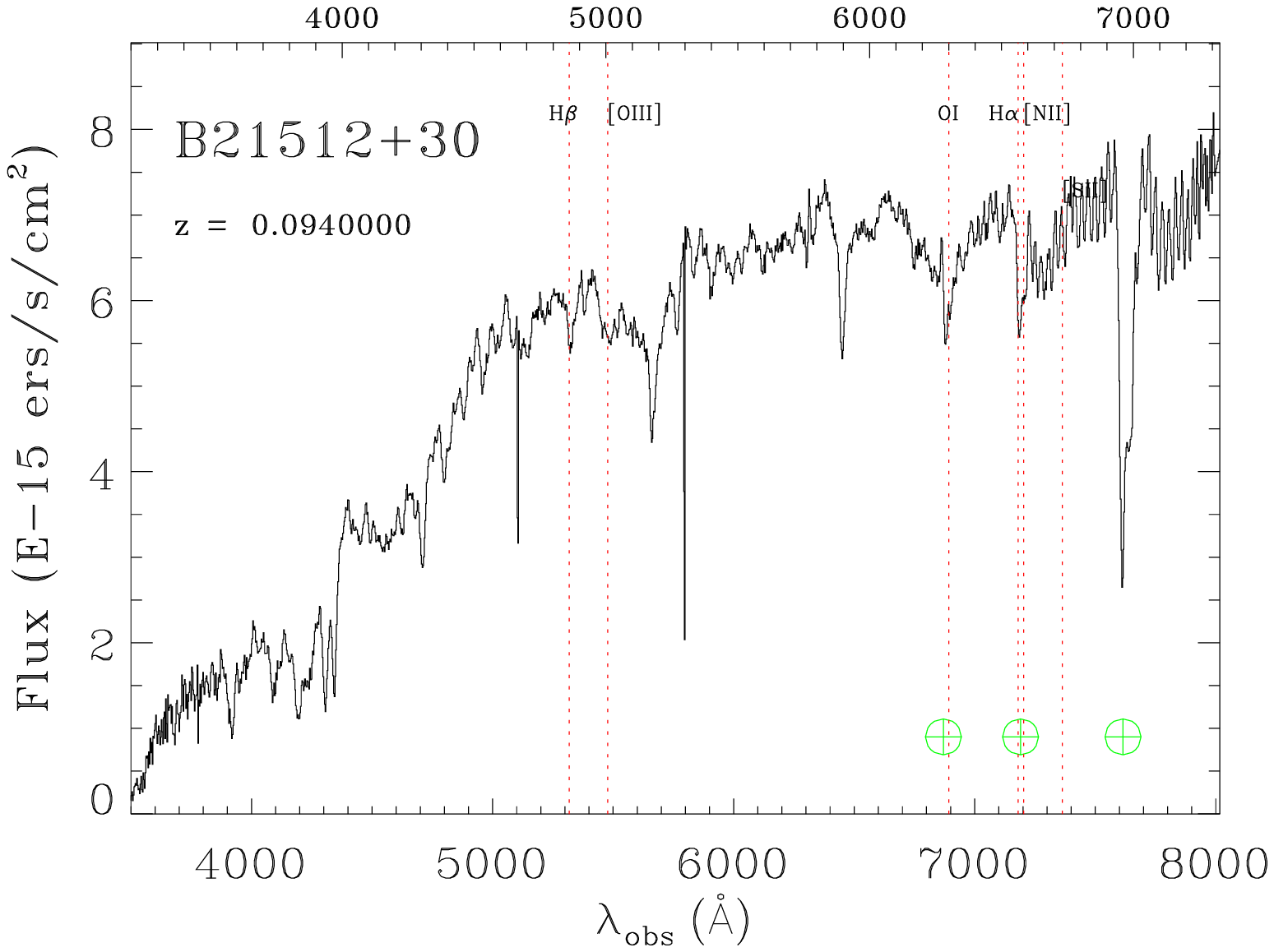}
\caption{{\bf Optical spectra of our TNG targets.} These spectra are extracted from
 the central 1.65 arcsec of the galaxies. The fluxes are in units of $10^{-15}$
  erg s$^{-1}$cm$^{-2}$.
The red dotted vertical lines indicate the position of the main
 optical emission lines, while the green crossed circles on the bottom
 of the spectra indicate the position of the main telluric absorption
 lines. On the upper left side of the spectra the source redshift is
 indicated below the target name.}\label{FigSpectra}
\end{figure*}

\begin{figure*}[h!] 
\centering
\includegraphics[width=0.45\textwidth]{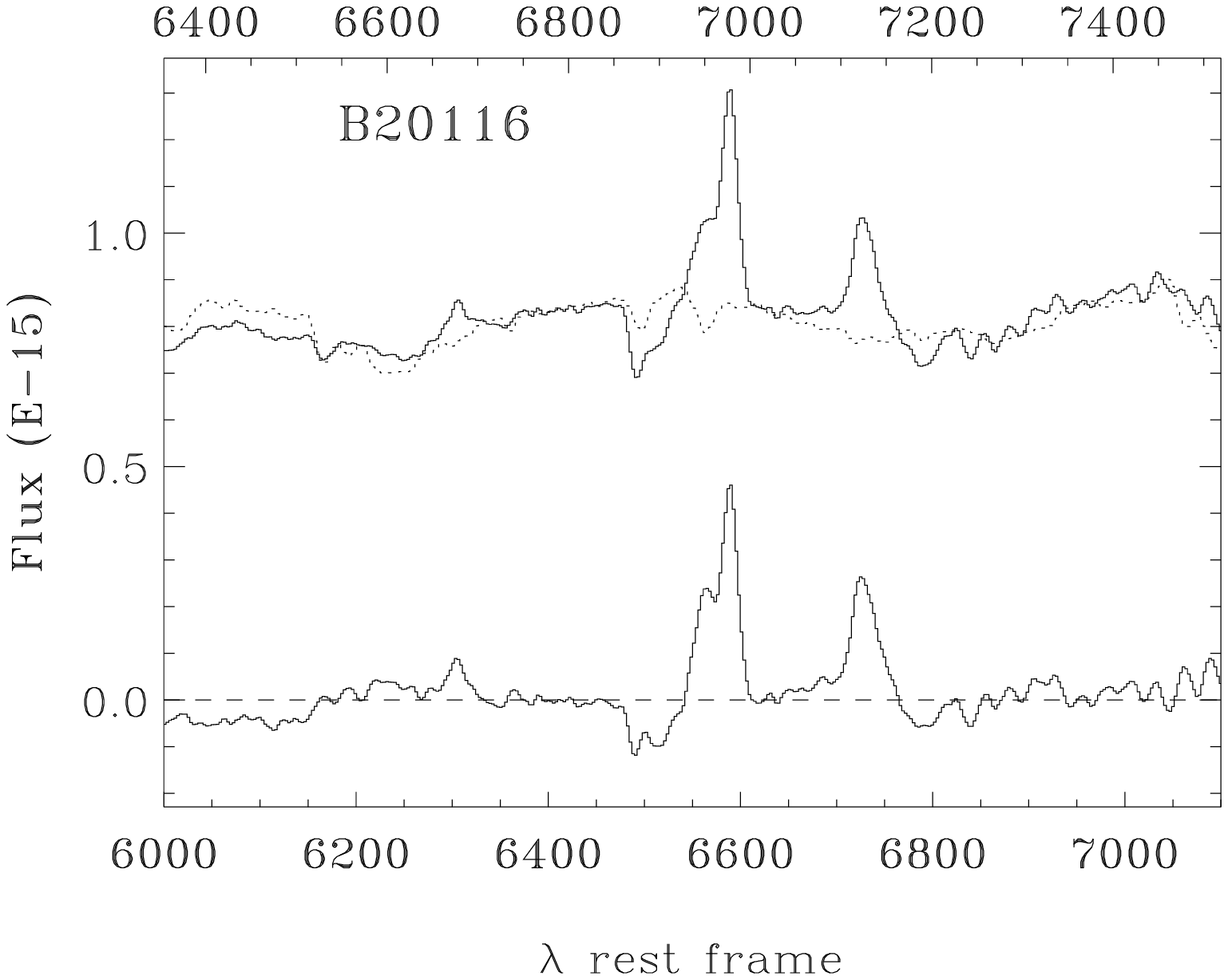}
\includegraphics[width=0.45\textwidth]{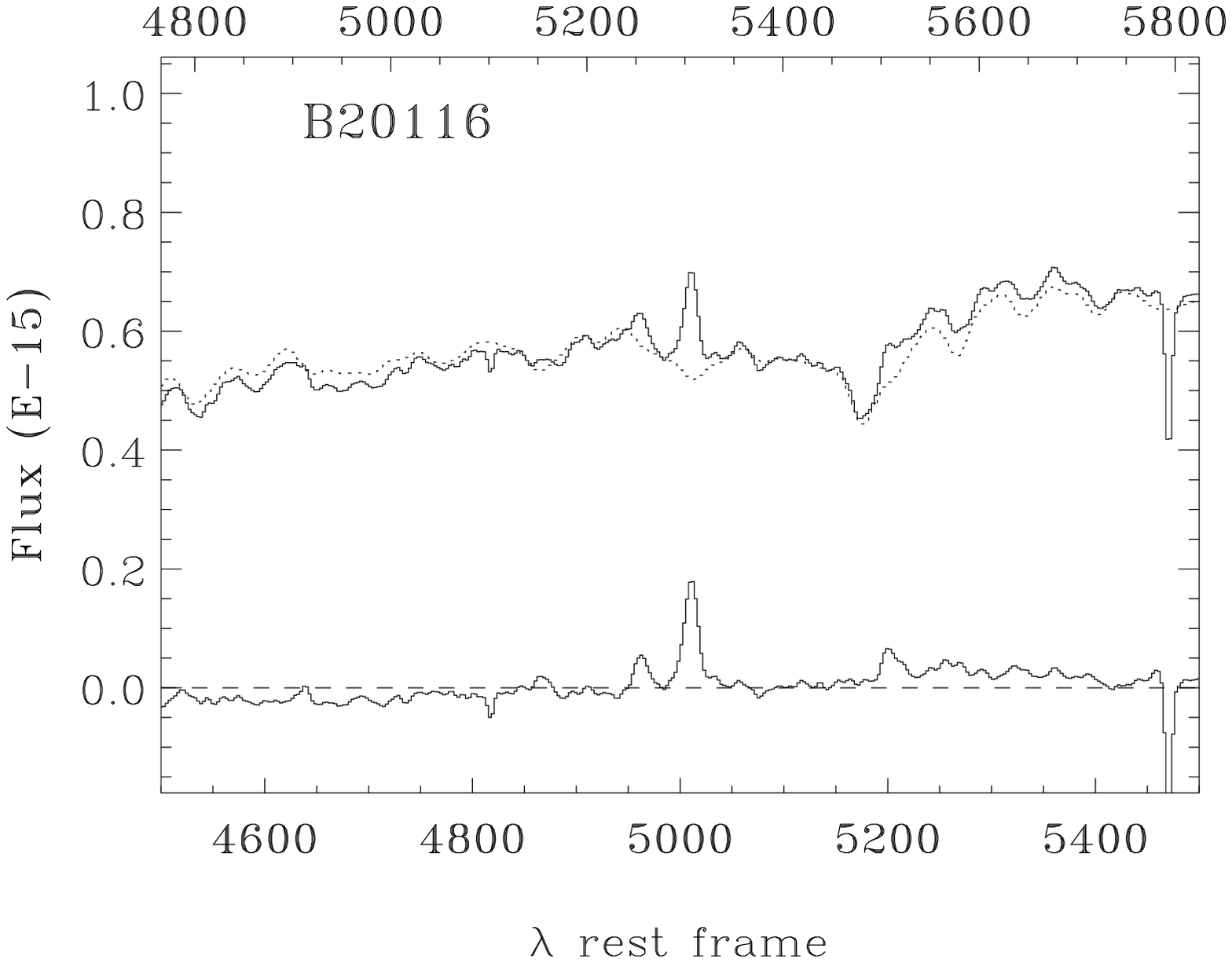}
\includegraphics[width=0.45\textwidth]{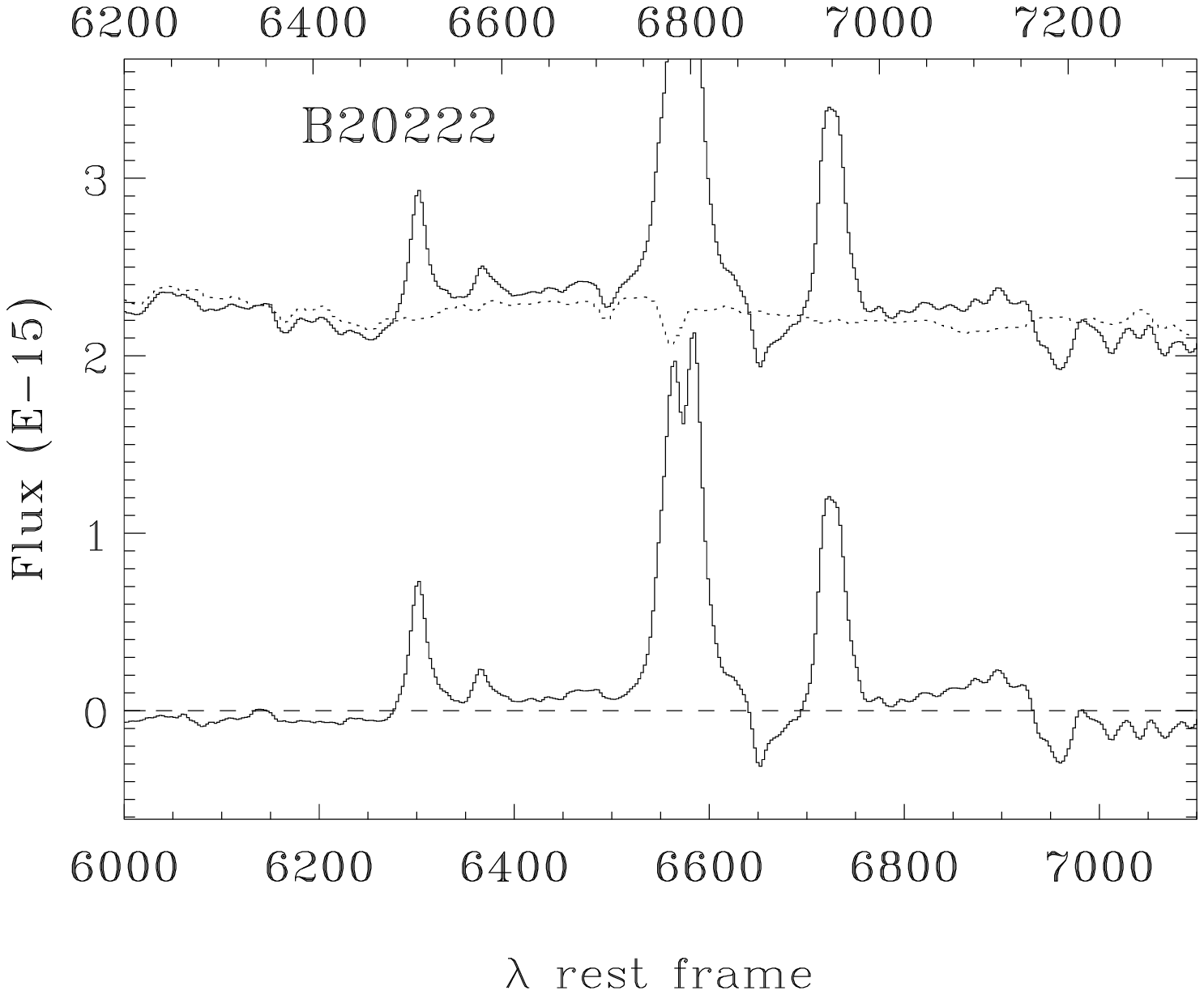}
\includegraphics[width=0.45\textwidth]{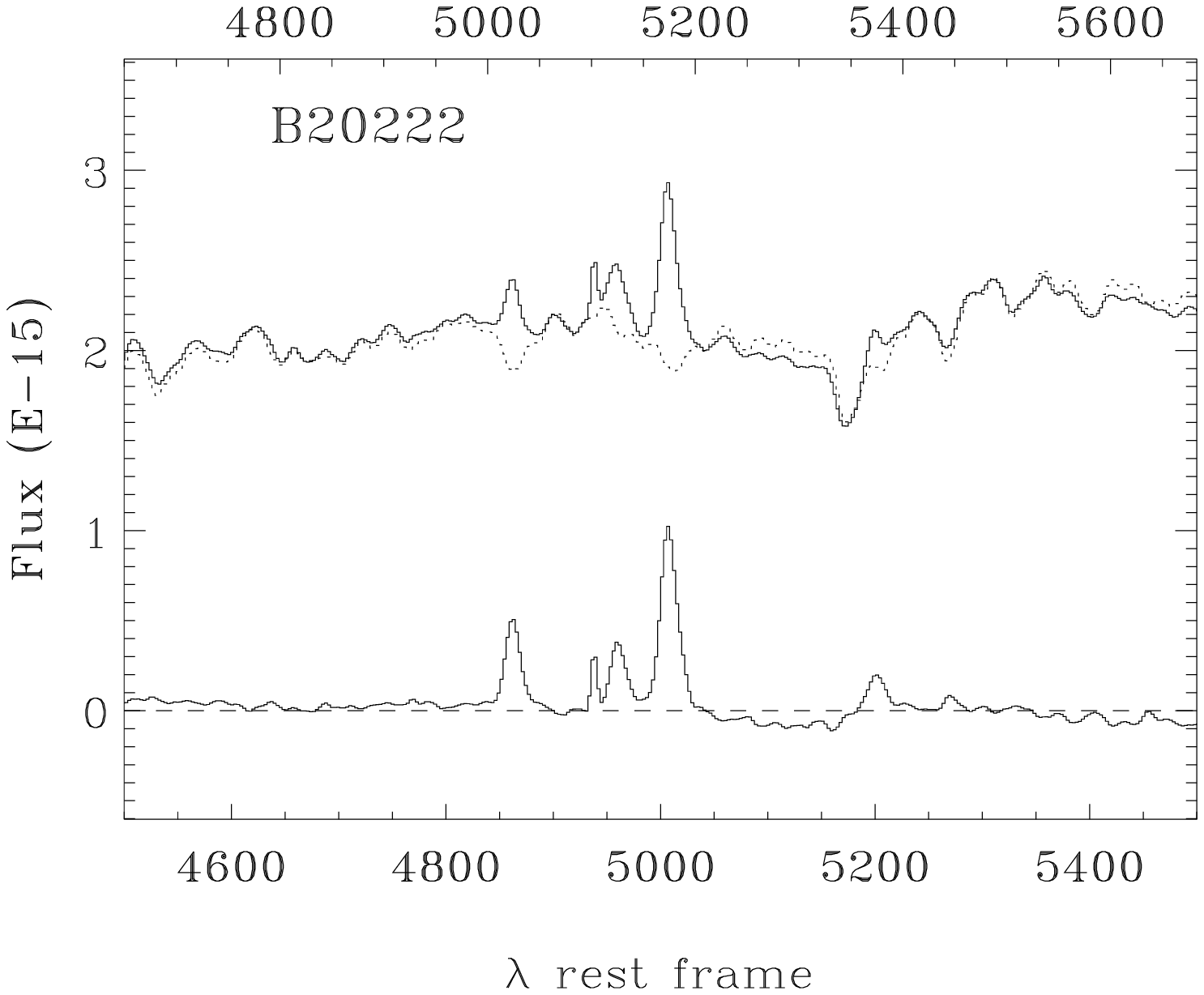}
\caption{ As an example of the adopted procedure, we show the 
{\bf spectra of two C BCSs} after the
  subtraction of the single stellar population model with the best fit. The 
first column represents
  the modeling centered on the H$\alpha$ line, while in the second column 
there are the fits for the H$\beta$ line. In all the pictures, the   C BCS 
spectra are in solid lines, the top spectra are before the stellar removal 
and the bottom spectra are the results of the host galaxy stellar population 
subtraction. The dotted line through the top spectra indicates
  the best fit single stellar population model. The dashed line across the bottom 
spectra indicates the zero flux level.}\label{FigStellarSub}
\end{figure*}


\section{OPTICAL RESULTS.}

\subsection{Measure of the emission lines luminosities}
\label{line}

In Tab.4 we report the emission lines fluxes.
For targets observed with the TNG, we measured the lines intensities using 
the {\it specfit} package to fit Gaussian
profiles to each emission line in our sources: H$\beta$, 
[OIII]$\lambda$$\lambda$ 4959.5007 \AA, [OI]$\lambda$$\lambda$ 6300.64 \AA, 
H$\alpha$, [NII]$\lambda$$\lambda$ 6548.84 \AA, [SII]$\lambda$$\lambda$ 
6716.31 \AA. To reduce the number of free parameters, we forced the velocity 
and FWHM to be the same for all the lines.
  The integrated fluxes of each line were free to vary except for those with 
known ratios from atomic physics: in this case the 
[OI]$\lambda$$\lambda$ 6300.64 \AA, [OIII]$\lambda$$\lambda$
4959.5007 \AA\ and [NII]$\lambda$$\lambda$ 6548.84 \AA\ doublets. Where 
required, we insert a flat continuum.
In some cases, we tried to fit a broad component for both the  H$\beta$ and 
H$\alpha$ lines, but the resulting lines intensities were smaller enough with 
respect to the narrow components to let us ignore them. 
For objects with information available in literature, we took emission line 
measurements as reported by the authors (see References in Tab.1).

\begin{table*}
\caption{{\bf New TNG emission line fluxes.} The sources are 
listed in
    the first column; in the following columns
the fluxes of the major emission lines are listed. The fluxes are in units
of $10^{-15}$ ergs$^{-1}$cm$^{-2}$. The 
upper limits to the line measurements are indicated by $<$. } 

\label{tab_emissionLine}
\centering
\begin{tabular} {llllllll} 
\hline
\hline
Name    & H$\alpha$ & [NII]$\lambda$6584 \AA & [OI]$\lambda$6364 \AA& [SII]$\lambda$6716 \AA &  [SII]$\lambda$6731 \AA& H$\beta$ & [OIII]$\lambda$5007 \AA\\
\hline \hline
B2 0116+31     &(3.9 $\pm$0.4)   &(9.2 $\pm$0.4)   &(2.5 $\pm$0.3)   &(5.4 $\pm$0.4)   &(2.5)   &(0.54$\pm$0.09)   &(3.05$\pm$0.11)   \\
B2 0222+36     &(3.3 $\pm$0.1)e+1&(4.0 $\pm$0.1)e+1&(1.7 $\pm$0.1)e+1&(2.1 $\pm$0.1)e+1&(1.7)e+1&(9.2 $\pm$0.4)   &(2.10$\pm$0.05)e+1\\
B2 0258+35A    &(3.1 $\pm$0.1)e+1&(6.4 $\pm$0.1)e+1&(1.8 $\pm$0.1)e+1&(3.2 $\pm$0.1)e+1&(1.7)e+1&(8.1 $\pm$0.9)   &(3.8 $\pm$0.1 )e+1\\
B2 0648+27     &(1.4 $\pm$0.1)e+4&(3.55$\pm$0.02)e+4&(1.9 $\pm$0.2)e+3&(2.3 $\pm$0.2)e+3&(1.9)e+3&(8.7 $\pm$0.8)e+3&(9.88$\pm$0.01)e+4\\
B2 1037+30     &(1.00$\pm$0.05)   &(1.7 $\pm$0.05)   &(3.2 $\pm$0.4)e-1&(8.4 $\pm$0.6)e-1&(3.2)e-1&(0.4 $\pm$0.2)e-1&(6.0 $\pm$0.3)e-1\\
B2 1855+37     & -                &$<$7.8e-1         & -            & -                & -      & -                & -                \\
\hline
B2 0844+31B    &(7.5 $\pm$1.5)e-1&(1.5 $\pm$0.2)   & -  & -    & -  & - &(2.1 $\pm$0.3)\\
B2 1003+35     &(6.5 $\pm$0.1)   &(4.66$\pm$0.08)   &(2.3 $\pm$0.1)   &(3.51$\pm$0.09)   &(2.33)  &(1.5 $\pm$0.1)   &(4.8 $\pm$0.1)   \\
B2 1512+30     & -                & -                & -                & -                & -      &$<$2.0 e-1        &(5.1 $\pm$0.9)e-1\\
\hline
\hline
\end{tabular}
\end{table*}

\subsection{Diagnostic diagrams.}
\label{diagr}

Diagnostic diagrams are constructed from pairs of observed line ratios which 
reveal information on ionizing continuum, ionization parameter, gas 
temperature and other physical properties of the emission line regions.
According to Heckman 1980, Baldwin et al. 1981, Kewley et al. 2006 and 
other works, star forming galaxies are separated from  AGNs and AGNs into High Excitation Galaxies (HEG) and Low Excitation Galaxies (LEG).These ratios are chosen so that considered lines are very close to avoid reddening and extinction problems (strongest for the bluer 
part of the spectrum), and also the ratios can be 
measured even if there are uncertainties on the flux calibration of the 
spectra.
Moreover, the chosen lines are often the strongest features of the optical 
spectra, easily found also in low luminosity galaxies.

In Fig. 3 we show the three diagnostic diagrams in Tab. 1:\\
([OIII]$\lambda5007 /$H${\beta}$) versus ([NII]$\lambda6583/$H${\alpha}$),([OIII]$\lambda{5007}/$H${\beta}$) versus ([SII]$\lambda{6716} \lambda{6731}/$H${\alpha}$) and the 
([OIII]$\lambda{5007}/$H${\beta}$) versus ([OI]$\lambda{6364}/$H${\alpha}$).
 The standard optical diagnostic diagram is the 
([OIII]$\lambda5007 /$H${\beta}$) versus ([NII]$\lambda6583/$H${\alpha}$). 
These line ratios are used to separate the star-forming galaxies from AGNs, 
since the [NII]$/$H${\alpha}$ is more sensitive to the presence of low level 
AGN than other lines due to its sensitivity to metallicities, while the [OIII]$/$H${\beta}$ 
line ratio is sensitive to the ionization parameter of the gas (the amount of ionization 
transported by the radiation moving through the gas).
The ([OIII]$\lambda{5007}/$H${\beta}$) versus 
([SII]$\lambda{6716} \lambda{6731}/$H${\alpha}$) and the 
([OIII]$\lambda{5007}/$H${\beta}$) versus 
([OI]$\lambda{6364}/$H${\alpha}$) diagnostic diagrams are more sensitive to 
the hardness of ionizating radiation field, dividing the AGNs into two 
branches: high ionization sources {\it (i.e. High Excitation Galaxies, HEG)} 
lie on the upper branch, low ionization sources 
{\it (i.e. Low Excitation Galaxies, LEG)} lie on the lower branch.

 In Fig. 3, B2 0708+32B, B2 0722+30, B2 1257+28, B2 1855+37, 
B2 0331+39, B2 0844+31, B2 1101+38, B2 1512+30 are not 
considered due to their upper limits or undetected lines in at least one of 
the diagnostic ratios.
The star forming, HEG and LEG regions are separated by solid lines according 
to Kewley et al. 2006. The region between the dashed line and the solid 
line in the  Log ([OIII]$\lambda5007 /$H${\beta}$) versus 
Log([NII]$\lambda6583/$H${\alpha}$) 
diagram indicates the region of composite 
galaxies, sources with both star-forming and nuclear activity. We also 
plotted in color the HEG (grey) and LEG (cyan) regions occupied by the radio-loud AGN with 
redshift $z<0.3$ taken from the 3CR Third Cambridge Catalog of Radio Sources) sample (Buttiglione et al. 2010). The red dotted lines represent the HEG/LEG separation derived for 3CR. The
3CR catalog of radio sources is characterized by unbiased selection criteria 
with respect to optical properties and orientation, and it spans a relatively 
wide range in redshift and radio power, covering the whole range of behavior 
of radio-loud AGN. BCS sources, with detected enough emission lines to make diagnostic diagrams (see Tab. 4), are shown as colored dots according to their classification. 
The position of C BCS sources on the diagnostic diagrams indicates that the 
majority of our sources belongs to the LEG group. The only exceptions are 
B2 0648+27  and 3C 305, located in the HEG region. We note also that in this 
region is present B2 1144+35 a BCS FRI radiogalaxy. We also labeled the 
source B2 0149+35, source in a low activity phase detailed discussed in Sect. 5.1.2.

\section{OPTICAL-RADIO CORRELATION}
\label{corr}

The correlation between AGN optical narrow emission line and radio power is 
verified since long time (e.g. Baum et al. 1989a, 1989b). This correlation is 
explained thinking of a common energy source for both the optical lines and 
radio emission: the isotropically emitted radiation from the active nucleus 
ionizes the optical emitting gas and the radio luminosity is determined by 
the properties of the central engine.

As shown by Morganti et al. 1997, the same optical-radio relation holds both
for extended sources and compact sources. They compared the [OII]$\lambda$
3727 \AA\ and [OIII]$\lambda$ 5007 \AA\ emission line luminosities of a sample
of CSS sources with the values found for extended sources of similar radio
power and redshift. They found a very intriguing result: in the correlation
between the [OII]$\lambda$ 3727 \AA\ - radio luminosities, both compact and
extended radio sources lie on the same linear correlation; instead, looking at
the [OIII]$\lambda$ 5007 \AA\ - radio luminosities, compact sources tend to be
at the lower side of the [OIII]$\lambda$ 5007 \AA\ luminosity.

In Figs. 4 and 5, we have done respectively the [OIII]$\lambda$ 5007 \AA\ -
408 MHz radio extended luminosities and [OIII]$\lambda$ 5007 \AA\ - 5 GHz
radio core luminosities plots. We have included C and E BCS sources with
available optical data, comparing our results with different samples. B2
0708+32B and B2 0331+39 are not included because we have no information on the
[O III] emission line flux.   As we used a combination of our TNG, SDSS and other telescopes spectra,
 a potential issue concerning our optical
  spectroscopic data is the difference in the spatial size of the associated spectral
  aperture. This is due to the different angular sizes of the aperture and
  also to the range of redshift covered by our sample. However, we verified
  that no link is present between the instrumental setup (or redshift) and the
  location of the various objects in Fig. 4 and 5.

We added samples of more powerful compact sources (orange crosses: Gelderman
et al. 1994, magenta plus: Morganti et al. 1997), in order to verify the
presence of an optical-radio correlation. Moreover, we superimposed the 3CR
LEG and HEG sources (grey and cyan small squares, Buttiglione et al. 2010) to
compare compact sources with extended ones. Finally, we plot also (green
diamonds) the Core Radio Galaxies (CoRG) sample of Baldi $\&$ Capetti 2010 to
compare our sample with these peculiar faint compact sources.

From Fig. 4, we note that C BCS sources show a lower 
total radio power
than the LEG objects discussed by Buttiglione et al. 2010 or that their optical
emission line is higher than expected from the radio power.
C BCS sources show a significantly higher optical luminosity
with respect to the best linear fit
obtained by Buttiglione et al. 2010 for HEG and LEG sources (see cyan and
green dashed lines). 
CoRGs studied by Baldi $\&$ Capetti 2010 follow the same trend of C BCS. It looks like that 
core radio galaxies and
C BCS have in general a low total radio power with respect to HEG, LEG and CSS
radio sources, but the line optical luminosity is a factor two higher
with respect to the correlation with the total radio power.

We compared also at the [OIII]$\lambda$ 5007 \AA\  luminosity with the 
5 GHz radio core luminosities for all samples discussed before
(Fig. 5). In this diagram CoRGs as well as C BCS 
sources are in between the HEG and LEG best linear fit. 
Our C BCS sources seems to be intermediate objects
between 3CR LEG + CSS sources and the CoRGs of Baldi $\&$ Capetti 2010.
Nuclear emission of 3CR HEG sources is definitely brighter in  [OIII] 
luminosity, while C BCS at a given core radio power are near to the linear
bets fit of LEG sources but on average C BCS are optically brighter.

We note that also CSS 
sources do not follow the two linear fits suggesting a possible common fit
of CSS, C BCS and Core radio galaxies. It seems that they follow an independent
track with respect to HEG and LEG sources. 
Kunert-Bajraszewska et al. 2010, discussing a sample of Low Luminosity
Compact Objects, suggest two parallel evolutionary tracks for HEG and LEG 
sources, evolving from GPS (Gigahertz Peaked Sources) to CSS to FR.
Our diagram, thanks to the addition of
low power Core Radio galaxies and C BCS sources selected at low frequency
and without any constrain on their spectral index, suggest a more complex 
scenario.

\begin{figure*}[t!]
\centering
\includegraphics[width=1\textwidth]{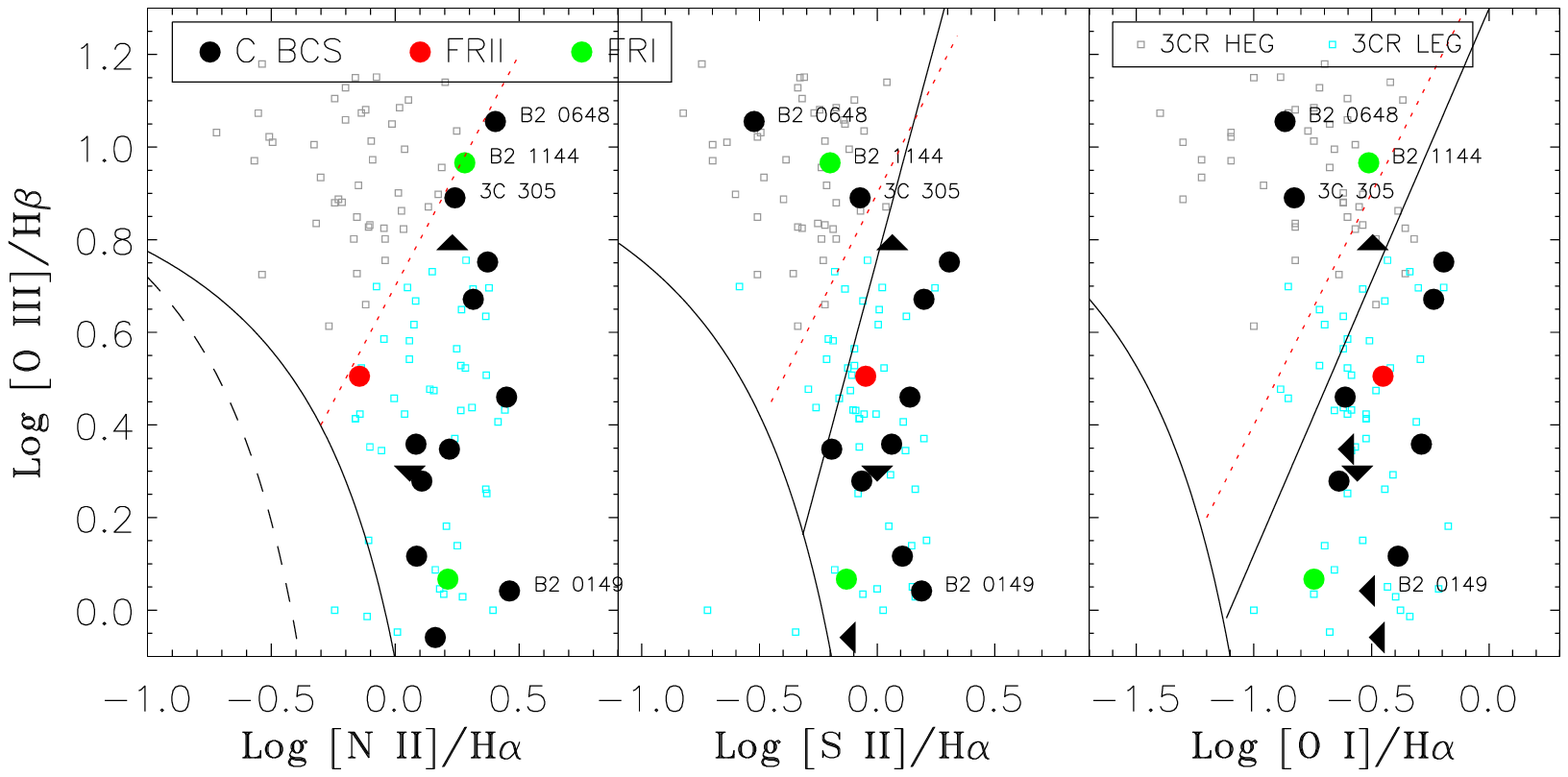} 
\caption{ {\bf Diagnostic diagrams}: The curves are taken from Kewley et al. 
2006 and divide galaxies into star forming, HEG and LEG AGNs. The region between the dashed line and the solid line in the first diagram indicates the position of composite galaxies. The dotted 
lines mark the empirical boundaries between HEG (grey squares) and LEG (cyan squares) in each diagram as derived for the 3CR sample (Buttiglione et al. 2010). The C BCS are indicated with black circles, FR II E BCS with 
red circles and FR I E BCS with green circles (see Tab. 5). Upper limits (UL) on the emission 
line measurements are indicated by triangles pointing towards them 
(i.e. an UL in H${\beta}$ is indicated with a  triangle pointing up.)}
\label{FigDiagnostic}
\end{figure*}

\clearpage

\begin{figure*}[t!]
\centering
\includegraphics[width=1\textwidth]{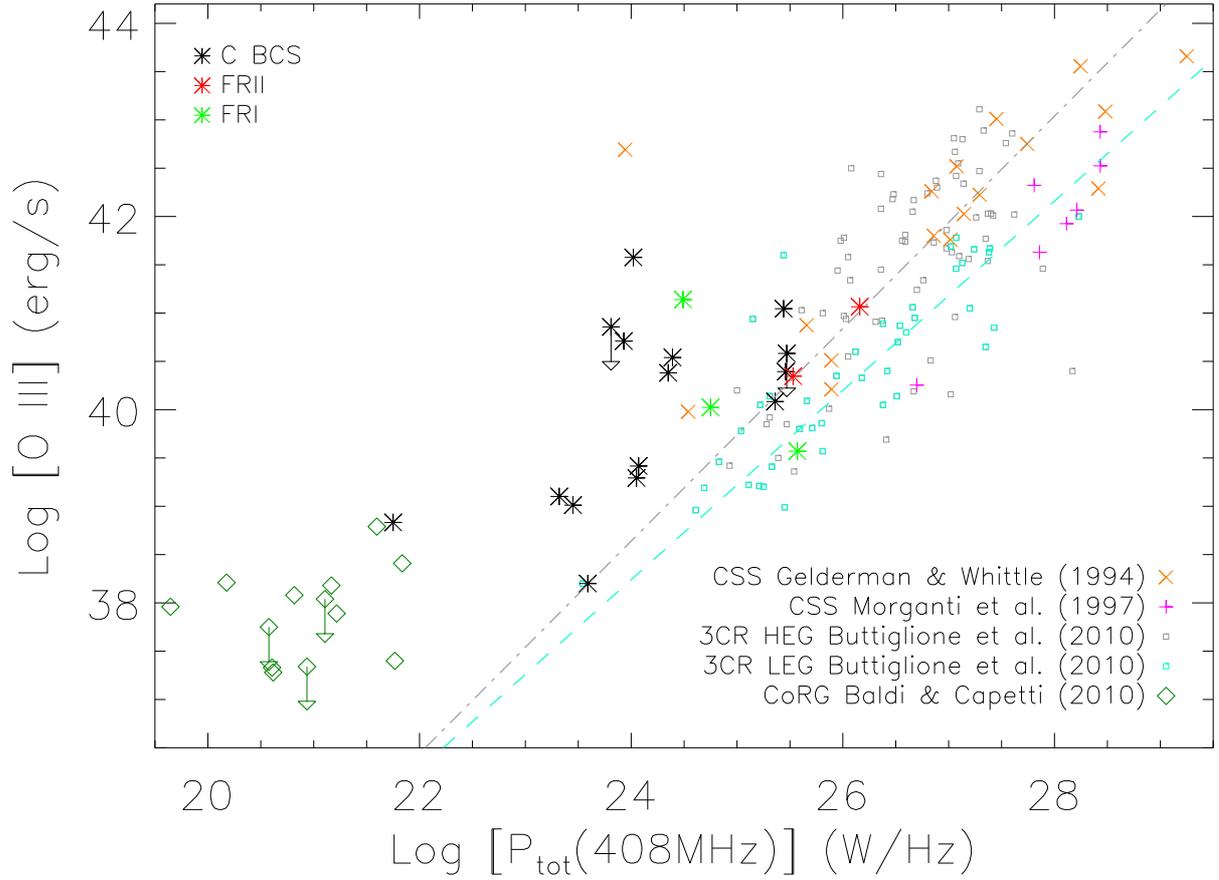} 
\caption{In this plot we present the 
correlation {\bf total radio power at 408 MHz vs the [O III] emission 
line intensities} produced by the narrow line regions. The two dashed lines 
represent the best linear fit obtained for the HEG and the LEG sub-populations
separately (according to Buttiglione et al. 2010). Many samples are here
compared: see text for more details.} \label{optradE}
\end{figure*}

\clearpage

\begin{figure*}[t!]
\centering
\includegraphics[width=1\textwidth]{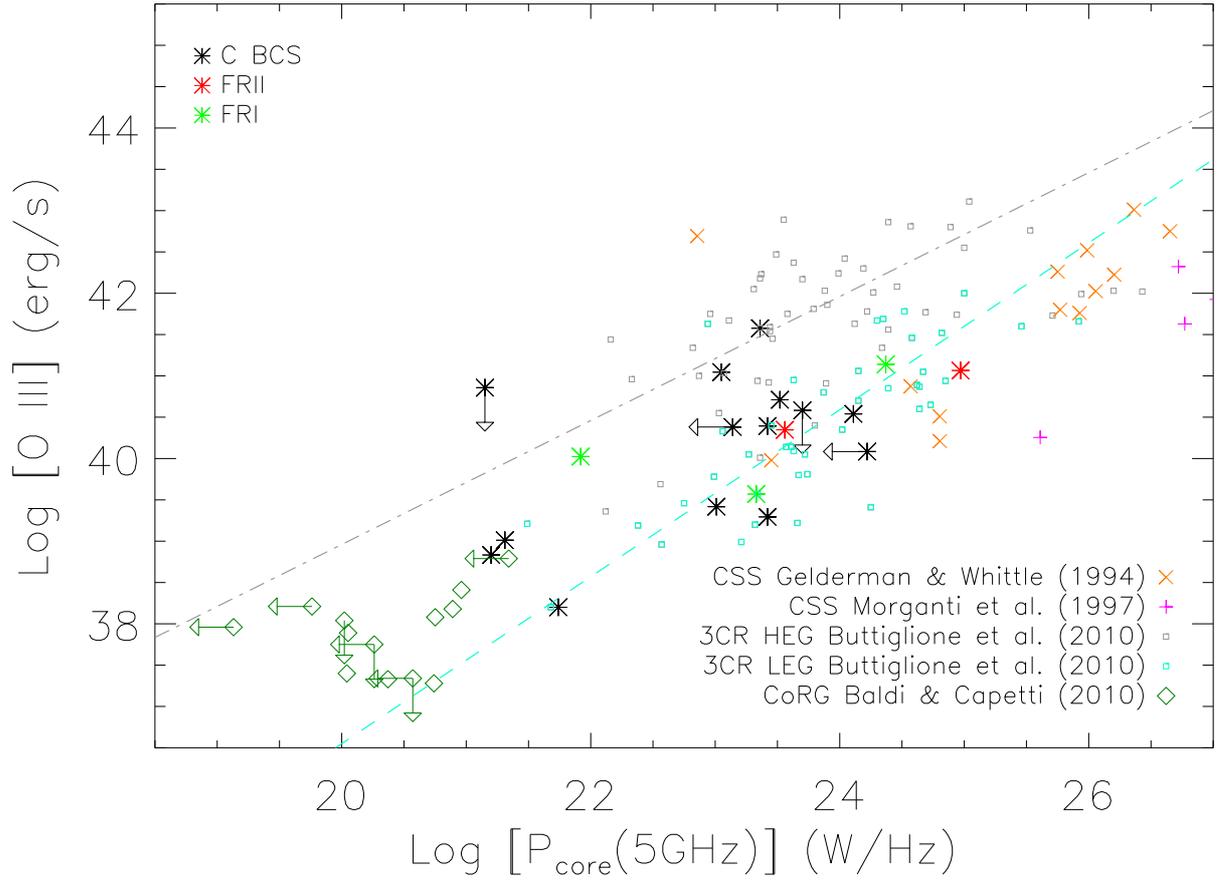} 
\caption{ {\bf [O III] luminosity [erg s$^{-1}$]  as a function of radio core 
power at 5 GHz in W Hz$^{-1}$}: the two dashed lines represent the best linear 
fit obtained for the HEG
and LEG sub-populations separately according to Buttiglione et al. 2010. 
Symbols and colors show different samples as in Fig. 4.}
\label{optradC}
\end{figure*}

\clearpage

\section{NOTES ON SINGLE SOURCES.} 

\subsection{Compact BCS sources} \label{notesC}

In this Section, we give short notes on the whole sample of C BCSs. For all 
resolved targets in our new radio maps, we present here also an image. For all 
these sources, if not specified, the beam size and the noise level of each 
map are those reported in Tab.2. 
In Tab.5, we give basic source parameters at 8.4 GHz and 22 GHz 
for our new VLA images. 
The reported core positions are obtained from our new VLA observations.

The arcsecond core flux density at 5 GHz and total flux density at 408 MHz 
are from Liuzzo et al. 2009b.

\begin{table*}[th!]
\caption{{\bf Sources parameters} from the new VLA observations.}
\begin{center}
\footnotesize
\label{tabSource}
\begin{tabular}{cccccccc}
\hline
\hline
       &           &           &                    &8.4 GHz    &22 GHz     &                      &\\
Source &Component  & RA (J2000)& DEC (J2000)        & flux      &flux       &$\alpha^{22}_{8.4}$   &size\\
       &           & h m sec   & d \arcmin~~~~ \arcsec  &(mJy)     &(mJy)       &                      &$^{\prime \prime}$\\
\hline 
\hline
B2 0149+35     &core      &01 52 46.458 &  36 09 06.50 &5.5  &4.9    &0.12   & P\\
\hline
B2 0708+32B    &core      &07 11 47.704  &  32 18 35.12 & 12.6    &7.5   &0.54     & P\\
            &N lobe    &-             &-              & 2.7    &-      &-       &3.5\\
            &S lobe    &-             &-               &3.2    &-      &-       &3.0\\
            &total     &-             &-               &18.5   &-   &-    &13.0\\
\hline
B2 0722+30     &core      &07 25 37.256  & 29 57 14.96   & 14.6     &6.7     &0.77   & P\\
            &E lobe    &-             &-             & 4.9      &-        &-      &8.0\\
            &total     &-             &-             &19.0       &6.7     & 1.08    &12.0\\
\hline
B2 1254+27     &core      &12 57 24.356  & 27 29 52.23  & 1.1    &5.3     &-1.63  & P\\
            &total     &-             &-              & 1.2   &5.6    &-1.60  & 0.3\\
\hline
B2 1557+26     &core      &15 59 51.616  & 25 56 26.35  & 13.5   &12.3    &0.10   & P\\
             &NE jet   &-              &-             & 1.8    &-       &-     &0.7\\
            &total     &-              &-             & 15.3   &12.3    &-   &0.6\\
\hline
\hline
B2 0331+39     &core      &03 34 18.415& 39 21 24.37& 145.8   &114.5  &0.25     &  P\\
            &S jet     &-             &-             &28.9     &13.9   &0.76    &0.9\\
            &halo      &-             & -            & 29.5    & -   &-       &3.5 \\
            &total     &-             &-             &204.2    &136.2  & -   &3.5\\
\hline
\hline
\end{tabular}
\end{center}
{\scriptsize In Col.8, P indicates point-like structure and it is related to the core morphology (see text in Sect. 2.1).}

\end{table*}

\subsubsection{B2 0116+31.}

The radio galaxy B2 0116+31 (4C 31.04) is 
classified as a low-redshift CSO (Compact Symmetric Object) (Giovannini et al. 2001, Cotton et al. 1995). VLBA (Very Long Baseline Array) images at 5 GHz show a compact core component with hot spots and lobes 
on either side (Giroletti et al. 2003). Spectral line VLBI (Very Long Baseline Interferometer) observations 
reveal the presence of an edge-on circumnstellar H I disk (Perlman et al. 2001). 
The optical nucleus is also found to have cone-like 
features well aligned with the radio axis. 

According to Perlman et al. 2001, optical data suggest a relatively recent 
merger having occurred  $\gtrsim 10^8$ yrs ago. In our TNG spectrum we detect 
all the diagnostic emission lines and their ratios clearly settle the source 
among LEG.

\subsubsection{B2 0149+35.} \label{0149}

B2 0149+35 is identified with NGC 708,
a D/cD galaxy associated with the brightest galaxy in the central 
cooling flow cluster Abell 262 (Braine $\&$ Dupraz 1994). B2 0149+35 is a low 
brightness galaxy whose nuclear regions are crossed by an irregular dust lane 
and dust patches 
(Capetti et al. 2000). 

The interaction between the cooling gas and the radio source is discussed by 
Blanton et al. 2004.
The comparison between the total and core 
radio power and between radio and X-ray images suggests that at present the 
radio core is in a low activity phase (Blanton et al. 2004, 
Liuzzo et al. 2010).\\ 
Our VLA observations show an unresolved component with a flux density of 
5.5 mJy at 8.4 GHz, 4.9 mJy at 22 GHz, and a flat spectrum 
($\alpha_{22}^{8.4}$ $\sim$0.12).
The lower flux density in 5 GHz VLBA maps can be due to a slightly inverted 
spectrum or to some extension lost in VLBA
images and/or because of variability.

From the diagnostic diagrams and the [O III]-radio plots the source is 
classified as LEG.

\subsubsection{B2 0222+36.}
This source shows an halo-core structure at arcsecond resolution 
(Fanti et al. 1986). Sub-arcsecond 8 GHz VLA map resolve the structure into a 
core and two components on either side, while at 22 GHz it shows an S-shaped 
morphology with a dominant core and two lobes (Giroletti et al. 2005b). 

In VLBI images at 1.6 GHz, B2 0222 + 36 is two sided, with jets emerging in 
opposite directions along the north-south axis. Since there is evidence of a 
change of the jet direction in the inner region, Giroletti et al. 2005b 
speculate that the jet orientation is rotating because of instabilities of 
the AGN. This could explain the small size of the source because an unstable 
jet did not allow the growth of a large scale radio galaxy. 
In this case the old round halo could be due to the diffusion of radio 
emission during the orbit of the inner structure. 

From the diagnostic diagrams, this source is classified as 
a LEG-type.

\subsubsection{B2 0258+35.}

This source was studied with VLA and Merlin + EVN (European VLBI Network)
by Sanghera et al. 1995, who classified it as a
CSS source. VLA data show a double structure with a separation of 
1.1 \arcsec. EVN  + MERLIN images reveal an extended plume-like feature at both 
ends of the source and a jet-like feature in between. Sub-arsecond VLA images 
at 8 and 22 GHz reveal the same structure. The source appears to strongly 
interact with the ISM (InterStellar Medium) as shown by the large bending of the arcsecond 
structure of the SE lobe and the presence of a surrounding low brightness 
extended structure in the VLA images. 
Large amount of extended HI-disk in the central region of the source is 
detected (Emonts et al. 2006a, 2006b).

From the diagnostic diagrams, it is a LEG-type source.

\subsubsection{B2 0648+27}

This object is only slightly extended at the lowest frequencies 
(Parma et al. 1986). It was resolved into a double source extended about 
1 $^{\prime \prime }$  with VLA observations at 8.4 GHz (Giroletti et al. 
2005b).  Morganti et al. 2003 detect a large amount of H I. 
The two radio lobes do not show any evidence for jet-like structure or the 
presence of hot spots. 
 
Giroletti et al. 2005b estimated a minimum age for this source of about 
1 Myr.  Probably the external lobe regions are much 
older, confirming that this source is confined, and it is expected to remain 
compact similar to NGC 4278  (Giroletti et al. 2005a). The connection between 
the small size of the radio  emission and the presence of a major merger in 
this galaxy about 10$^9$ years ago, with the presence of a large amount of 
H~I in this galaxy (Morganti et al. 2003) is remarkable.

From the optical spectrum, we observe very strong high excitation emission 
lines and the Balmer lines in 
absorption, as the spectra produced by a dominant young stars population.
In the diagnostic diagrams, this source occupies the region of HEG galaxies.
 This is the most powerful source in optical band in 
our sample.

\subsubsection{B2 0708+32B.}

This source reveals an extended morphology in direction NS (Fanti et al. 
1986) with two symmetric lobes about 10 $^{\prime \prime }$ in size 
($\sim$10 kpc). In VLBA images at 5 GHz (Liuzzo et al. 2009b), a double 
structure oriented along P.A. (Position Angle) $\sim$ 150$^\circ $ and extended $\sim$4 mas is 
observed. Since no spectral index information is available, the 
identification of the core is not possible. 

Our 8.4 GHz map (Fig.6) shows an unresolved nucleus 
(Tab. 5) and two symmetric lobes in 
agreement with the structure present in the 1.4 GHz observations.
No jet-like structure is visible. In our 22 GHz VLA observation, only the 
core is visible (Tab. 5). 
The angular resolution of VLA images is too low to resolve the double 
structure visible in VLBA images. 
Using VLA data at 1.4 GHz from Fanti et al. 1986, we estimated the spectral 
index for lobes and central component. We derived $\alpha_{lobes}\sim$ 1 for 
the two lobes and for the central component: $\alpha_{1.4}^{8.4}$ = 0.3 
and $\alpha_{8.4}^{22}$ = 0.54.  The steepening of the high frequency 
spectrum and the high degree of symmetry of the source suggests that the 
compact central component could be identified as Compact Symmetric Source 
(CSO) and this small size radio source is a restarted young source.
We want also emphasize how rare it is to see extended emission in a CSO. Up 
to now the only known with this properties are 0108+388 (Stanghellini et al. 
2005) and 0402+379 (Zhang et al. 2003).  
We have no optical spectroscopic information on this sources, so its classification is not possible.

\begin{figure*}[th!] 
\centering
\includegraphics[width=0.45\textwidth]{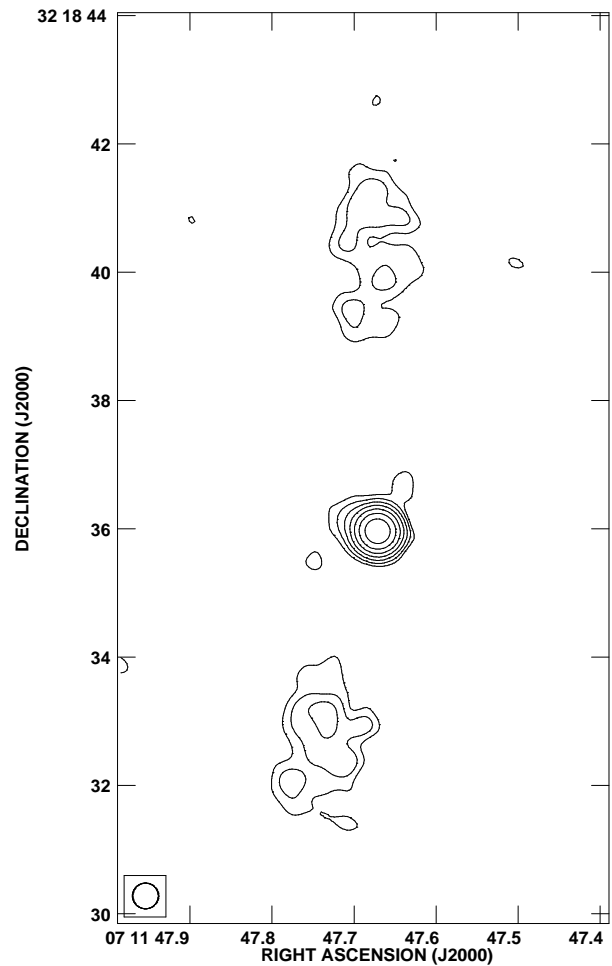}
\caption{{\bf 8.4 GHz VLA image of B2 0708+32B.} Contour levels are 0.1, 0.2, 
0.4, 0.8, 1.6, 3.2, 6.4 and 12.8 mJy/beam. The peak level is 11.8 mJy/beam. 
The beam size is 0.4 \arcsec $\times$ 0.4 \arcsec}\label{0708_8}
\end{figure*}

\subsubsection{B2 0722+30.} \label{0722}

This radio source is associated with a disk galaxy. Strong absorption is 
associated with its disk and a bulge-like component is also clearly visible. 
The radio emission originates from two symmetric lobe-like features in E-W 
direction, i.e. at an angle of $\sim$ 45$^{\circ}$ to the disk (Capetti et 
al. 2000). In Fig. 7, we report the optical image from the HST (Hubble Space Telescope) overimposed to a VLA radio image at 5 GHz.
 
At arcsecond resolution, this source shows an FRI radio morphology with a 
total power of 3.1$\times$ 10$^{23}$W Hz$^{-1}$ at 408 MHz and a linear size 
of $\sim$ 13.5 kpc.

\begin{figure*}[t!] 
\centering
\includegraphics[width=\textwidth]{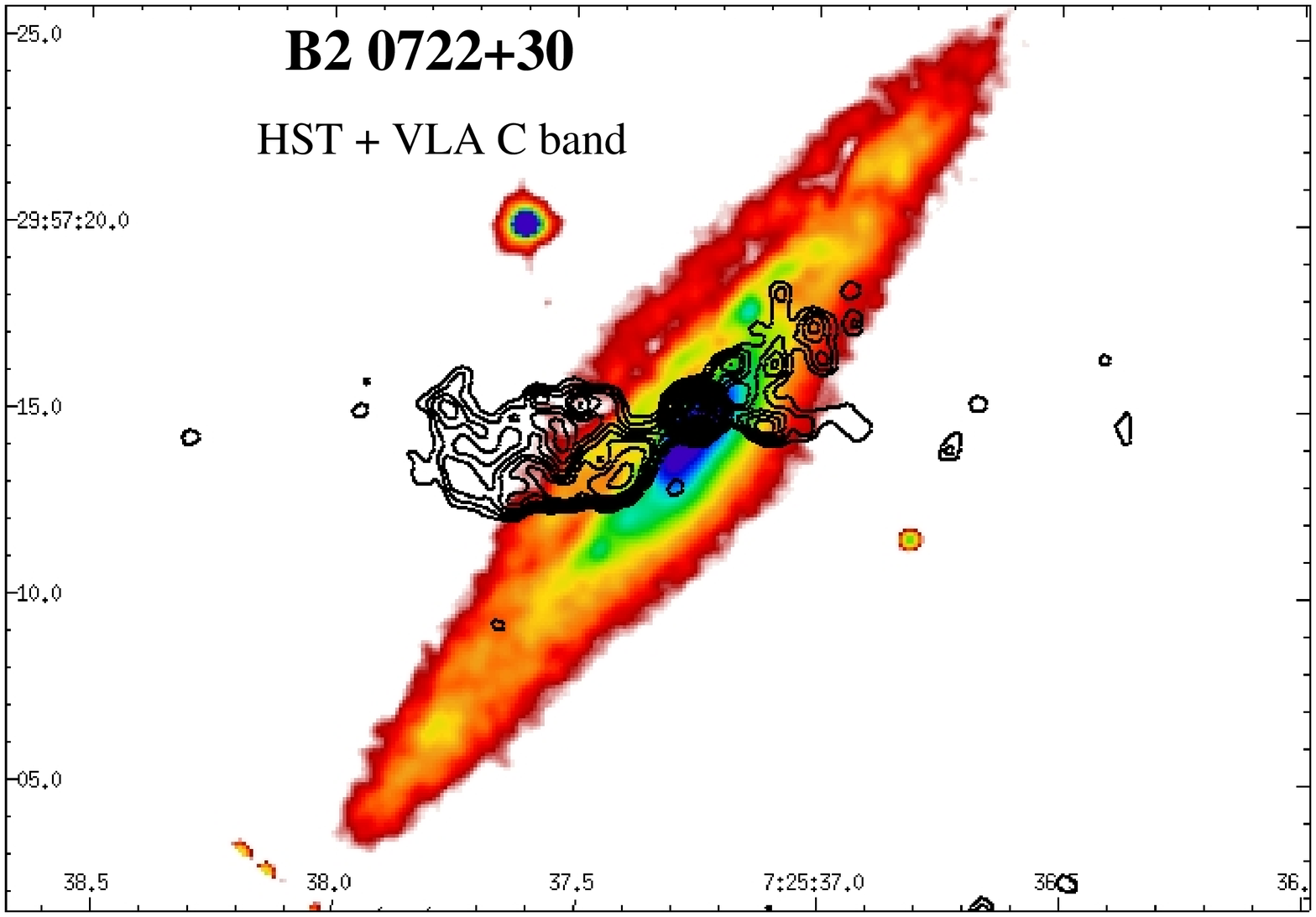}
\caption{{\bf VLA contour map in C band of B2 0722+30 overimposed to HST image.} Contour level for the radio map are 0.15, 0.20, 0.25, 0.30, 0.35, 0.4, 0.5, 0.7, 1, 2, 3, 5, 7, 9, 12, 15 mJy/beam. The beam size is 0.6 \arcsec $\times$ 0.6 \arcsec,0$^{\circ}$. The peak level is 20.6 mJy/beam.}\label{0722_hst+radio}
\end{figure*}

Our 8.4 GHz VLA observations (Fig. 8) show an unresolved nucleus, 
and a marginal detection of the East lobe which appears completely resolved. 
At 22 GHz, only the core is detected. It is unresolved with a total flux of 
6.5 mJy (Tab. 5). The radio core position is slightly offset 
with respect to the brightest optical region ($\sim$ 1\arcsec), however we 
note that the optical core is not well defined because of the large dust 
absorption.

The core radio spectrum shows a clear steepening: $\alpha^{1.4}_{4.8}$ = 
0.2, $\alpha^{4.8}_{8.4}$ = 0.64, and $\alpha^{8.4}_{22}$ = 0.85 suggesting 
a sub-arcsecond resolved structure in agreement with the not detection in our 
VLBA data (Liuzzo et al. 2009b).
  
\begin{figure*}[t] 
\centering
\includegraphics[width=0.5\textwidth, angle= -90]{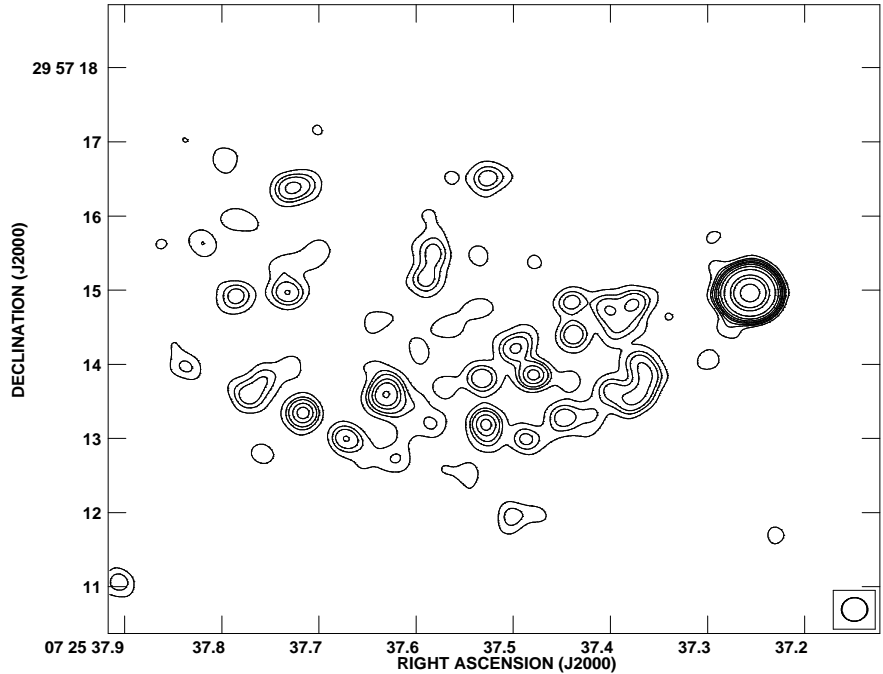}
\caption{{\bf 8.4 GHz VLA image of B2 0722+30.} Contour levels are  0.035, 0.07, 
0.1, 0.15, 0.2, 0.3, 0.5, 1, 3, 5, 10 and 15 mJy/beam. The peak level is 14.6 mJy/beam. }\label{0722_8}
\end{figure*}

This source is one of the rare objects where a disk galaxy 
in the nearby Universe hosts a radio emission FR I like shape. 
Literature research can provide information on only other three cases: 
1) the spiral galaxy 0313-192 in Abell 428 exhibiting a giant ($\sim$350 kpc) 
double-lobed FRI (Ledlow et al. 1998, 2001; Keel et al. 2006); 2) NGC 612 is 
a typical S0 galaxy with a large-scale star-forming H I disc and a powerful 
FR-I/FR-II radio source (V\'eron-Cetty $\&$ Ve\'ron 2001; Emonts et al. 2008); 
3) the BL Lac object PKS 1413+135 (McHardy et al.1994; Lamer et al. 1999). 

In the optical, we do not found either spectrum or the [O III] emission
line for this source: an optical classification is so not possible.

\subsubsection{B2 1037+30}

The source is only slightly resolved at 1.4 GHz (Fanti et al. 1986). At 
subarsecond resolution, at 8 GHz an edge-brightened structure, with complex 
sub-structures: jets, and lobes with hot spots is detected, while at 22 GHz 
only a point-like component, probably the core, and the resolved NW hot spot 
are evident. In VLBI images, the core is clearly revealed with a faint, 
diffuse emission detected on the shortest baselines (Giroletti et al. 2005b). 
According to its radio properties Giroletti et al. 2005b classified it
as a young CSO source.

In the optical, B2 1037+30 is identified with the brightest galaxy in the 
cluster Abell 923. From analysis, it follows the LEG sources: the optical line ratios set
the source LEG region close to the boundary with HEG sources. The
same position is found in the optical-extended radio plot, while in
the optical-core radio plot and in the accretion rate plot the source
in clearly identified with LEG types.
The optical line ratios set the source close to the HEG-LEG boundary. The same 
position is found in the optical-radio plots.

\subsubsection{B2 1101+38}

 Mrk 421 (B2 1101+38) 
is a well-known BL Lac, widely 
studied at all frequencies and detected at TeV energies (Punch et al. 1992). 
The NVSS (NRAO VLA Sky Survey) image shows a 30\arcsec core-dominated source, with emission on 
either side. 
The estimated viewing angle is $\theta \leq 20 ^{\circ}$. For more details on 
radio structure of this BL Lac, see Giroletti et al. 2006.

\subsubsection{B2 1217+29}

This source is identified with the nearby galaxy NGC 4278. 
In the radio band reveals a compact structure at all frequencies between 
1.4 and 43 GHz ( Condon et al. 1998; Di Matteo et al. 2001: Nagar et al. 2000, 2001).
Only at 8.4 GHz, at a resolution of $\sim$200 mas, 
it is slightly resolved into a two-sided source with an extension to the 
south (Wilkinson et al. 1998).
VLBA data show a two-sided emission on sub-parsec scales in the form of twin 
jets emerging from a central compact component (Giroletti et al. 2005a). 
Large amount of extended HI disc is detected in its central region (Emonts et 
al. 2006a, 2006b).

This source shows low [O III]/H$\beta$ ratio, settling in the LEG region 
of the diagnostic diagrams. In particular, it shows low [O III] luminosity, as 
confirmed by its position in the lower left side of the [O III]- radio plots.

\subsubsection{B2 1257+28}

This is one of two dominant members of the Coma cluster (Abell 1656) and considered the BCG. It is a cD galaxy whit a small size WAT (Wide Angle Tail) structure. Arcsecond scale properties are discussed in Feretti \& Giovannini 1987: it has a total flux density at 1.4 GHz of 190 mJy and the core flux density at 6 cm is 1.1 mJy. The radio emission is completely embedded in the optical galaxy, and a gap of radio emission is present between the core and the SW lobe, while a faint jet connecting the core and the NE lobe is detected. Parsec scale properties are discussed in details in Liuzzo et al. 2010: the source shows a one-sided structure with a total flux density of 10.1 mJy, and the core flux density is 7.27 mJy. 

Our optical analysis classified this source as LEG type object.

\subsubsection{3C 272.1}
3C 272.1 is identified with a giant elliptical galaxy. It shows strong 
radio emission and a two-sided jet emerging from its compact core. HST images 
reveal dust lanes in the core of the galaxy while no significant amount
of diffusely distributed cold dust was detected at sub-millimeter wavelengths 
(Leeuw et al. 2000).  At parsec scale, this nearby source shows a clear 
one-sided structure (Giovannini et al. 2001).

This source has low [O III] luminosity, as confirmed by its position 
in the lower left side of the [O III]- radio plots.

\subsubsection{B2 1254+27} \label{1254}
This radio galaxy is identified with the BCG (Brightest Cluster Galaxy) of a sub-group merging into the 
Coma  cluster. The optical galaxy, NGC 4839, is classified 
as a cD galaxy (Parma et al. 1986, and references therein).

In VLA observations at 1.4 GHz, it appears as a FRI radio source with 2 lobes 
in direction N-S. At 5 GHz, the arcsecond core flux density is $\sim$ 1.5 
mJy (Giovannini et al. 2005).

Our VLA maps at 8.4 GHz and at 22 GHz show a marginally resolved structure 
with an inverted spectral index (see Tab. 5), suggesting the 
presence of a compact nuclear emission.

The LEG classification of this source derived from the diagnostic diagrams is confirmed also 
by the [O III]-radio plots.

\subsubsection{B2 1322+36B}

On VLA scales, this source shows a twin-jet morphology. On parsec scale, it 
has a core-jet structure extending $\sim$ 10 mas in the same direction as the 
main large-scale jet. 

According to its position in all optical plots, this source is
classified as LEG.

\subsubsection{B2 1346+26}

This galaxy is the central galaxy of the cool core cluster A1795, which hosts 
the bright FR I radio source 4C 26.42. We studied in details its parsec scale morphology in Liuzzo et al. 2009a.
Our multi-frequency and multi-epoch VLBA observations reveal a complex, 
reflection-symmetric morphology over a scale of a few mas. The source appears two-sided with a well defined and symmetric Z-structure at $\sim$5 mas from the core. The kiloparsec-scale morphology is similar to the parsec-scale structure, but reversed in P.A., with symmetric 90$^{\circ }$ bends at about 2 arcsec from the nuclear region. A strong 
interaction with the ISM can explain the spectral index distribution and the 
presence of sub-relativistic jets on the parsec scale. 

In the optical, even if we detect upper limits on the [O III] emission line, the
position of this source in diagnostic diagrams reveals a LEG nature.

\subsubsection{3C 305}

This small FR I radio galaxy shows a plumed double structure with two faint 
hot spots and symmetric jets. The optical galaxy is peculiar, with continuum 
emission on the HST scale perpendicular to the radio 
jet (Giovannini et al. 2005, and references therein). The X-ray emission of this source is extended but 
it is not connected with radio structure. However, the X-ray emission is cospatial with the 
optical emission line region dominated by the [O III]5007. This could be 
interpreted as due to the interaction between the radio jet and the ISM 
(Massaro et al. 2009).

From our optical analysis, according to its position in Fig. 3, this source is
classified as HEG.

\subsubsection{B2 1557+26}

The host galaxy IC 4587 is a smooth and regular elliptical galaxy (Capetti et 
al. 2000).

In our 8.4 GHz VLA observations (Fig. 9), the radiosource is 
resolved in a core and a NE jet aligned with the emission observed in the 5 
GHz VLBA images by Giovannini et al. 2005. 
The jet is extended about 0.6 arcsec from the core.
At 22 GHz, the jet is not detected and the core has a total flux of 12.3 mJy 
(Tab. 5).

According to its position in all the optical plots, this source is
classified as LEG.

\begin{figure*}[t]
\centering
\includegraphics[width=0.5\textwidth, angle = -90]{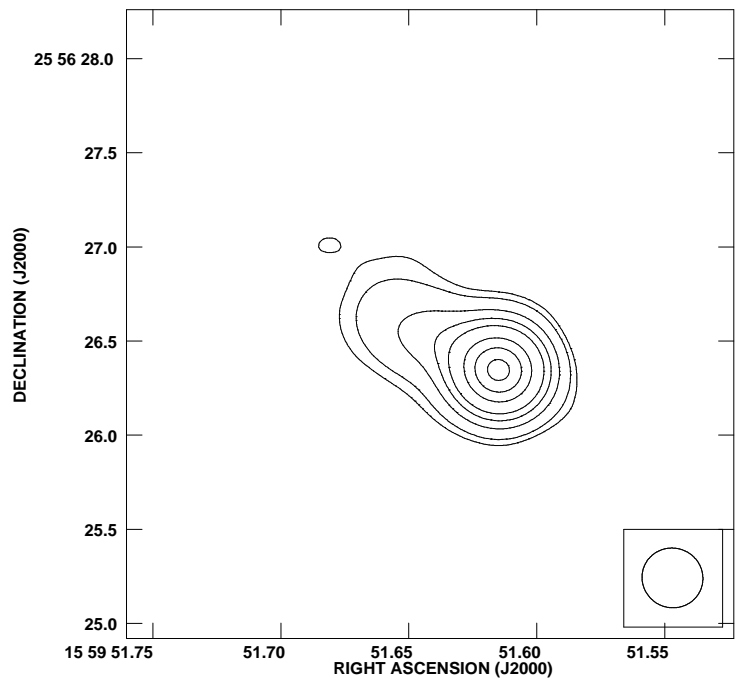} 
\caption{{\bf 8.4 GHz VLA map of B2 1557+26.} Contour levels are 0.15, 
0.3, 0.9, 1.8, 3, 6 ,9 and 2 mJy/beam. The peak level is 14.6 mJy/beam.}
\end{figure*}

\subsubsection{B2 1855+37}

This source shows a distorted double structure on the kiloparsec scale, with 
no detection on VLBA scale, that suggests its identification with a small 
symmetric source with a faint core (Giovannini et al. 2005, and references therein). The extended 
structure of this source seems to be confined by external gas pressure.

Due to upper limits on the optical emission lines, an optical
classification is not possible.

\subsection{Extended C BCS sources.} \label{notesE}

\subsubsection{B2 0331+39.}

Previous high resolution VLA observations at 1.4 GHz (A and B configuration) of 
this source (4C 39.12) revealed a resolved core plus a 
faint halo $\sim$1 arcmin in diameter (Parma et al. 1986, Fanti et al, 1986). 
Data at 5 GHz also show that the core is resolved and the inner source 
region is dominated by a bright one-sided structure extended 
$\sim$ 1 \arcsec, and surrounded by a symmetric low brightness halo.
 
This structure is confirmed by our 8.4 GHz VLA image 
(Fig. 10): the radio source is characterized by a nuclear 
emission plus one sided jet in South direction and a halo around it.
We note that the bright jet extension is not limited by sensitivity, but it 
looks like that the bright jet really ends after $\sim$ 1 \arcsec.

Our 22 GHz VLA map (Fig. 10) shows an unresolved core and the 
Southern jet. However, because of the high angular resolution and low surface 
brightness at this high frequency, the extended halo is not visible (see 
Tab.5). 

From Parma et al. 1986, we derived the spectral index between 1.4 GHz and 
8.4 GHz of the halo: $\alpha_{8.4}^{1.4}\sim$0.4. 
We produced also a spectral index image of the arcsecond jet comparing 8.4 
and 22 GHz images obtained using the same uv-range and angular resolution 
(Fig. 11). 
The core region has a flat spectral index with a clear steepening up to 
$\sim$ 0.9 along the jet, which appears steeper than the surrounding halo.

The one-sided jet structure is in good agreement with previous VLBA 
observations of this source presented in Giovannini et al. 2001 where the 
parsec scale jet is found at the same P.A. of the arcsecond jet presented 
here. 

At lower resolution this source is highly peculiar. From the NVSS image 
(see Fig. 10), we see a bright one-sided structure with a size 
of $\sim$3 arcmin (70 kpc) oriented at P.A.= 220$^\circ$,
i.e. very different from the jet and inner halo P.A. (160$^\circ$). 
Moreover, a fainter symmetric diffuse emission is present on a larger angular 
scale (more than 6 arcmin, 150kpc) at about the same P.A. of the inner 
one-sided jet and halo. The total NVSS flux density is 1.1 Jy. 
The total spectral index is 0.5 -- 0.6 from 74 MHz up to 4.8 GHz (see NED 
archive data), with a clear evidence that the new restarted component is 
dominant at all frequencies.

This complex morphology suggests a restarting activity with a change of the 
P.A. in the different epochs. A more detailed study is necessary to 
understand this peculiar source.

We do not found either the optical spectrum or the [O III] emission
line for this source: an optical classification is not possible.

\subsubsection{B2 0844+31}
This is a symmetric narrow-line FR II radio galaxy. At parsec resolution a 
bright core and two-sided jets are visible. The image from the FIRST 
Survey shows two extended FR I lobes at a larger distance from the core 
beyond the FR II-type hot spot, indicating a prior phase of radio activity. 
For this reason, this source could be classified as a restarted source, i.e., 
a source in which the extended FR I structure is related to previous activity,
whereas the inner FR II structure originates from more recent core activity 
(Giovannini et al. 2005). 

According to its position in all the optical plots, this source is
classified as LEG.

\subsubsection{B2 1003+35}
It is the largest known FRII radio galaxy with its projected linear size of 
more than 6 Mpc. It shows a complex structure at arcsecond resolution with 
evidence of relativistic jets oriented in the same direction as the 
large-scale structure with some oscillations (Giovannini et al. 2005). Its 
peculiar radio structure has been interpreted as evidence of restarted 
activity (O'Dea et al. 2001).

According to its position in all the optical plots, this source is
classified as LEG type, but close to the HEG boundaries.
 
\subsubsection{B2 1144+35}

This is a large scale ($\sim$ 0.9 Mpc) FRI radio source, core dominated with a 
short and bright two-sided jet. 
The bright arcsecond scale core is resolved at milliarcsecond resolution into 
a nuclear source, a main jet with an apparent superluminal velocity, and a 
faint counter-jet. Evidences of a dynamic interaction with the surrounding 
medium are present. The radio morphology of this source shows clear 
discontinuities at different linear scales suggesting a continuous activity 
but with high and low level periods (Giovannini et al. 2007).

From the optical point of view, this galaxy falls among HEG sources in the diagnostic diagrams, showing a very strong
[O III]$\lambda$$\lambda$ 4959,5007  \AA  doublet. This classification
is confirmed with the optical line luminosity - total and core radio
power plots, as well as in the accretion rate plot.

\subsubsection{B2 1512+30.} 

The host galaxy does not show any outstanding morphological features, except 
for very faint elongated dust absorption at the center (Capetti et al. 2000). 
In VLA image at 1.4 GHz with 5 arcsec of resolution, it appears as double 
source with two lobes in direction NS (Fanti et al. 1987), in 
agreement with the FIRST (Faint Images of Radio Sky at Twenty-Centimeters) image.
The angular size is $\sim$ 22 arcsec corresponding to $\sim$ 38 kpc.

Both in our 8.4 GHz and 22 GHz images, the source is undetected above 
$\geq$ 0.25 mJy/beam and $\geq$ 0.30 mJy/beam in X and K bands respectively.\\
Using the total flux at 408 MHz and at 1.4 GHz, we derived a spectral index 
$\alpha_{1.4GHz}^{408 MHz}$$\sim$ 2.
The non detection of a core emission at high frequency together with the steep low frequency spectral index suggest that 
this object is a dying radiosource with a radio quiet core.

From our optical study, according to its position in the diagnostic diagrams (Fig. 3), this source is
classified as LEG.

\subsubsection{B2 1626+39}

This source is identified with the central galaxy  3C 338 in the cool core cluster 
A2199. It is a 
multiple nuclei cD galaxy with the presence of dust lanes (Jensen et al. 2001).
On kiloparsec scales it has two symmetric extended radio lobes, characterized 
by a steep spectrum and misaligned with the central emission. The two radio 
lobes are connected by a bright filamentary structure.  Both the steep radio 
spectrum and strong filamentary emission may be caused by interactions with 
the dense intracluster medium (Gentile et al. 2007).  3C 338 was the first 
radio source in which a two-sided jet was observed on parsec scales 
(Feretti et al. 1993).

In the optical, according to its position in the optical-radio plots, this source is
classified as LEG source.

\begin{figure*} [th!]
\centering
\includegraphics[width=1\textwidth]{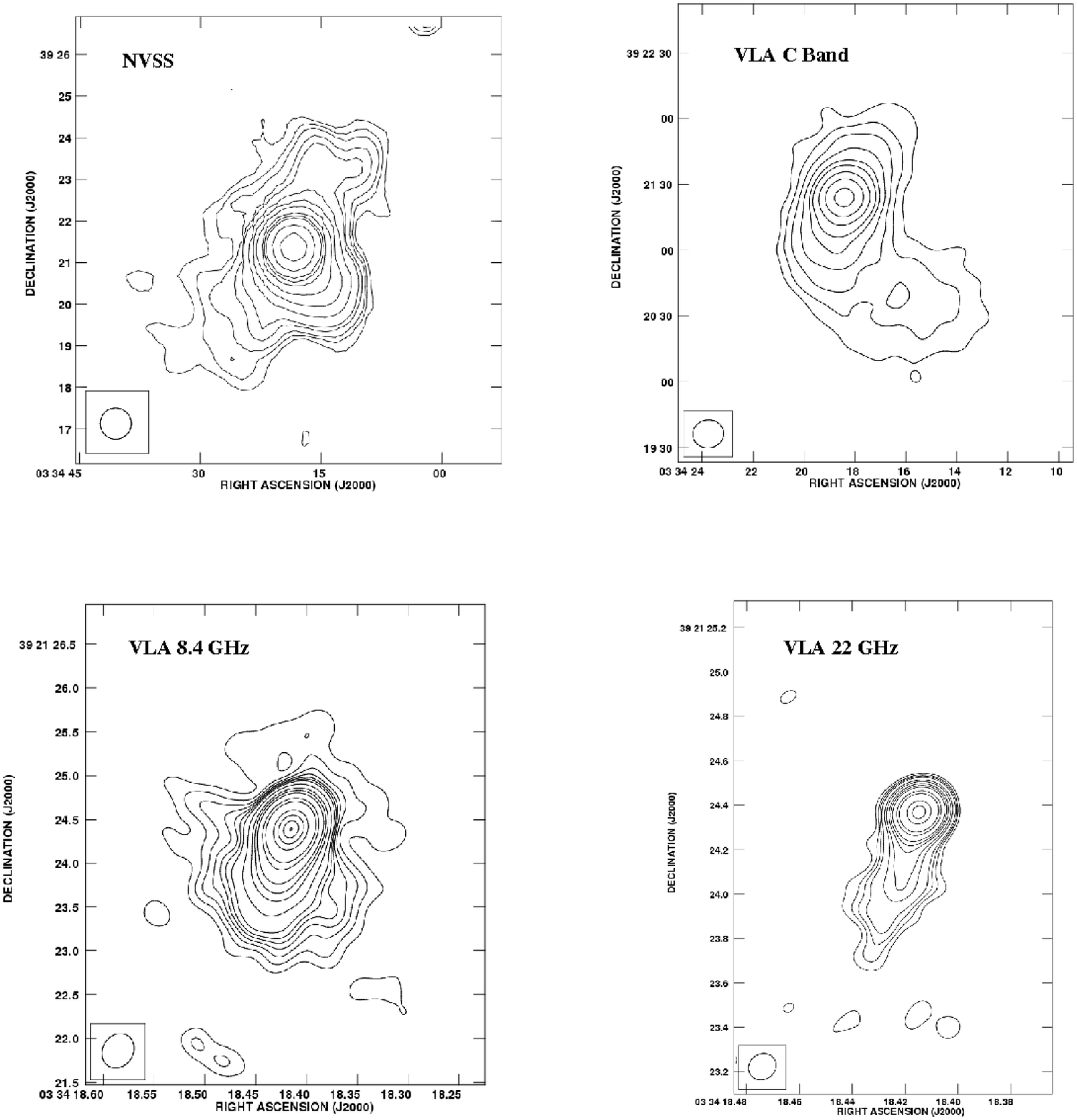} 
\caption{Top left: {\bf NVSS image of B2 0331+39}. Contour levels are 1, 1.5, 2, 3, 4, 8, 16, 32, 64, 90, 128, 256 and 512 mJy/b. Peak level is 811.2 mJy/beam. Beam size is 45 \arcsec $\times$ 45 \arcsec, 0$^{\circ}$. Top right: VLA radio image in {\bf C Band} of B2 0331. Contour levels are 1.6, 3.8, 7.8, 15.6, 31.1, 51.2, 107.9, 218 and 436 mJy/b. Peak level is 556.2 mJy/beam. Beam size is 14 \arcsec $\times$ 12.7 \arcsec, -83$^{\circ}$. Bottom left: {\bf 8.4 GHz VLA image} of B2 0331+39. Contour levels are 0.7, 0.9, 1.1, 
1.3, 1.5, 1.7, 2, 3, 5, 7, 10, 20, 30, 50, 70, 100, 130, 150 mJy/beam. The beam size is 0.41 \arcsec $\times$ 0.35 \arcsec, P.A.= -33. The noise level is 0.1 mJy/beam. Bottom right: {\bf 22 GHz VLA image} of B2 0331+39. Contour levels are 0.3, 0.5, 0.7, 1, 2, 3, 5, 7, 10, 20, 30, 50, 70, 100 mJy/beam. The peak level is 117.1 mJy/beam. }
\label{0331_allres}.
\end{figure*}

\clearpage

\begin{figure*} [h!]
\centering
\includegraphics[width=0.60\textwidth]{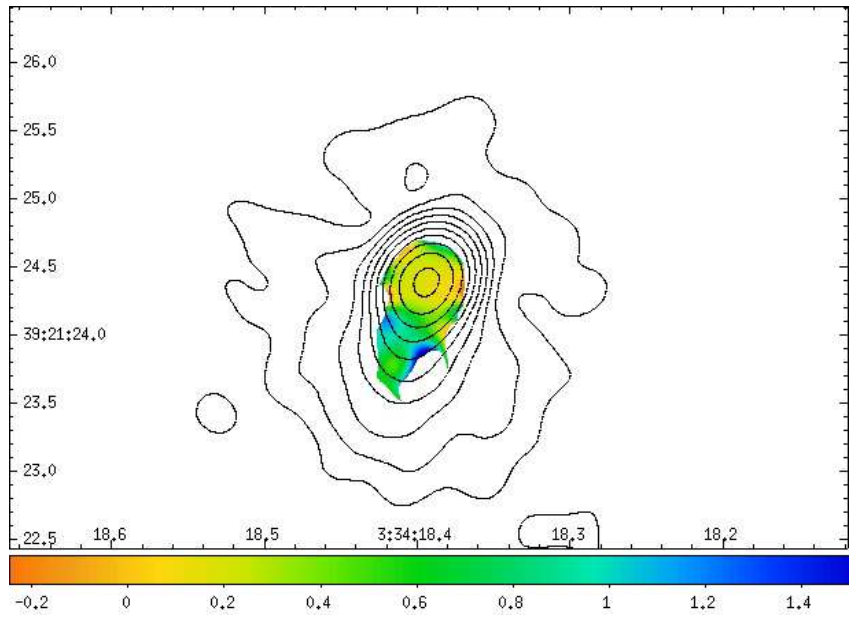}
\caption{{\bf Spectral index map (color) of B2 0331+39} between 8.4 and 22 GHz 
with overimposed the 8.4 GHz VLA emission (contour image).} 
\label{0331_spix}
\end{figure*}

\clearpage

\section{DISCUSSION.} \label{discussion}

We show in Fig. 12 a radio power vs 
linear size diagram
for the 95 sources of the whole BCS sample.
Looking at the radio properties of C BCSs (red squares), some compact sources are in 
agreement with a general correlation between the linear size and the radio 
power, and they show a radio power at 408 MHz lower than 10$^{24}$ W/Hz.
However, about half of compact sources show a radio power 
larger than 10$^{24}$ W/Hz .i.e. are in the same range of classical 
extended radio galaxies. 

\begin{figure*}[th!]
\centering
\includegraphics[width=0.8\textwidth]{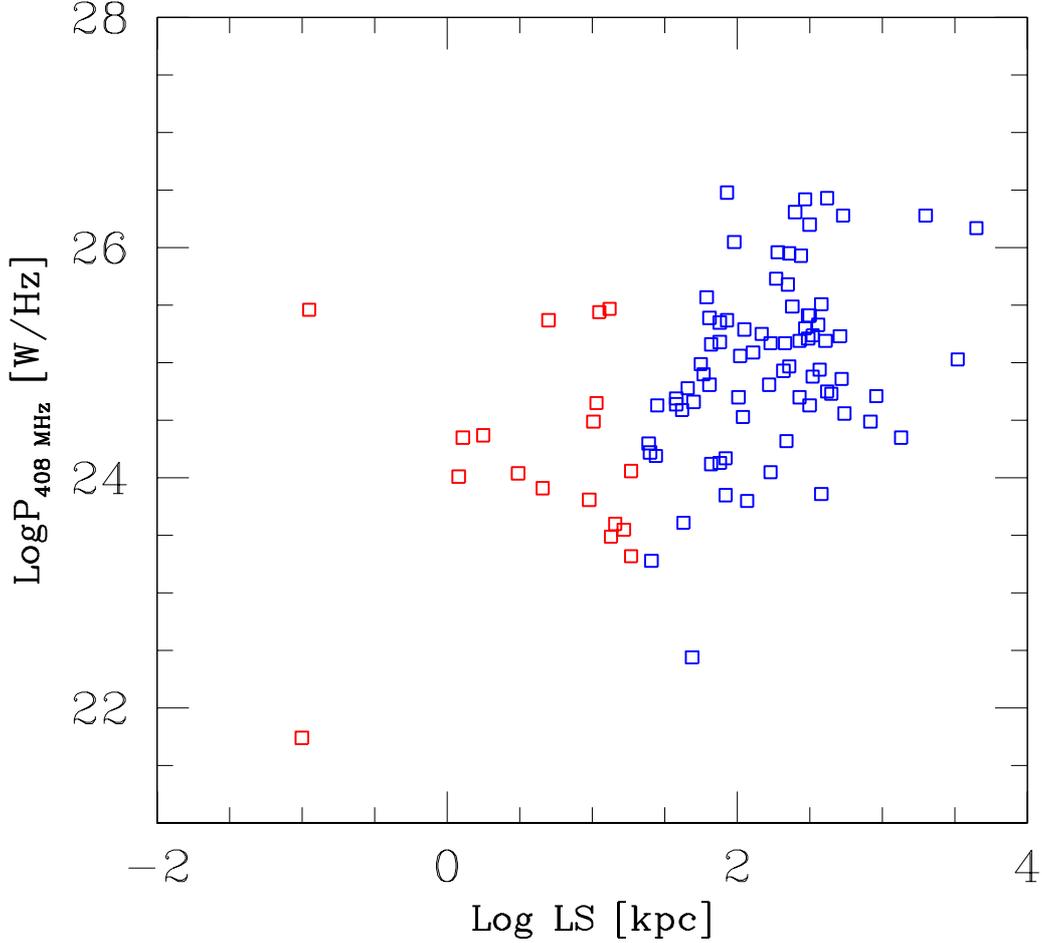} 
\caption{{\bf Radio power vs. linear size} diagram for sources in the Bologna Complete Sample (BCS, Giovannini et al. 2005). The  red squares indicate C BCS sources , while the blue ones represent the remaining extended objects.}\label{plot_PL}
\end{figure*}

To better investigate the nature and properties of C BCS properties we 
compared optical and radio data.
From Fig. 3, we note at first that the majority of C BCS 
sources are LEG. Only a couple of exceptions are present. 
These peculiar compact sources are B2 0648+27 identified with a high rich HI 
galaxy (Emonts et al. 2006a, 2006b), and 3C 305 identified with a peculiar gas rich
galaxy discussed in detail by Massaro et al. 2009. We note that also one FR I
BCS is among HEG galaxies. This source studied in details by Giovannini et al.
2005 show evidence of a multi phase radio activity. Recently, it restarted the
radio activity and it shows on the parsec scale structures moving with
an apparent velocity larger than c.
We conclude that for all the 3 cases, the optical type could be explained by 
the peculiar activity of the source inner region (restarted activity or strong
confinement and interaction with the surrounding medium). 

From Fig. 4, we point out that C BCS sources show a correlation between total radio power and [OIII] Luminosity
in agreement with Core Radio galaxies of Baldi $\&$ Capetti 2010 and not with FR I radio galaxies.
Also the three HEG sources discussed just before are in the same correlation
while the few other BCS FR I or FR II sources show the same correlation of 3C 
radio galaxies.
These considerations seem to suggest us that among LEG galaxies two different populations are present
according to their radio and [OIII] luminosity: FR I radio galaxies and
Core radio galaxies + C BCS. 
In particular, we note that, at given total radio power, compact sources have an
overluminous [O III] emission at a given radio power with respect to
HEG and LEG 3CR sources. Moreover, if we consider the core radio power at 5 GHz, compact 
sources show a different trend with respect to LEG radio galaxies: they are
overluminous in optical at low radio power (CoRGs) and with 
properties similar to 3CR FRI at higher radio power (C BCS). Finally, data
suggest that CSS could follow the
same trend of compact sources, becoming optically underluminous at 
very high radio power. A larger statistics is necessary to clarify this point.

To better understand this correlation we take the opportunity of many
data available for C BCS sources to investigate their origin.

Among C BCSs,  we have quite different sources:
\begin{itemize}

\item
{\bf a source can be compact because of projection effects:} 

In our sample only one source is clearly oriented at a small angle with 
respect to the line of sight: B2 1101+38 (Mkn 421, Tab. 6). This source is a well 
known BL-Lac type object (Giroletti et al.2006) and its size is affected by 
strong projection effects. We note, that as discussed in Liuzzo et al. 2009b, 
the percentage of sources oriented at small angle with respect to the line of 
sight in the BCS is in agreement with unified models prediction.\\

\item
{\bf a source can be compact because young or restarted activity:} 

In our sample many sources show evidence of recent nuclear activity as CSO
and CSS sources (e.g. B2 0258+35, Tab. 6), or restarted/recurrent activity (e.g. B2 0149+35, Tab. 6). These sources are expected to have a strong
interaction with the dense inner ISM (see e.g. B2 1346+26, the BCG in A1795, Tab. 6).
Peculiar source is B2 1855+37 which is characterized by a low-no jet emission could be in the 
final stage of this scenario, and we do not know if it will die or will
restart a new radio activity.If we compare these type of compact sources with the extended BCSs with evidence of a restarted (e.g. B2 0331+39, B2 0844+31, B2 1144+35, Tab.6) or dying activity (e.g. B2 1512+30, Tab.6) we do not found any differences in their OIII luminosity and 408 MHz total power.

We note also that on 18 C BCSs, 6 sources are in clusters/groups and they are the BCGs. 
Literature studies (O'Dea et al. 2001) observed that compact, low power
and steep spectrum radio sources in 
BCGs are not rare. Radio properties of these objects could be explained if 
nuclear fueling is related to the AGN activity cycle and we can see 
galaxies in a period of relative AGN quiescence, or just restarted (young).  
The presence of restarted and cyclic radio activity in such clusters are 
requested by the cooling scenario (McNamara et al. 2005, and references therein).  
We claim that the fraction of C BCS sources that are BGCs is considerable, corresponding, as discussed above, to the 1/3 (6/18) of the C BCSs. If we consider the whole BCS sample, the fraction of BGCs is 10/95. Among the 10 BCGs, the majority of them (6/10) are compact BCS sources. This seems to suggest that the strong interaction with the dense ISM of the cluster environment increases the source probability to be compact radiosource, due to frustration effect and/or jet instability. However, optical properties of extended BCGs (see Tab. 6 and Sect. 5) and compact BCGs are similar. \\

\begin{table*} 
\caption{ {\bf Properties of Compact and Extended BCS sources.}}\label{macr}
\begin{center}
\begin{tabular}{cccccccc}
\hline
\hline
Name   & Morphology & Optical&LLS & logP(408) &  Log$L_{[OIII]}$& Type& Notes\\
  &  & clas &kpc & W/Hz& (erg/s)&  & \\
\hline
\hline
B2 0116+31 & CSO   &LEG& 0.1 & 25.46 &  40.39 &  young & 4C31.04\\
B2 0149+35 & FR I  &LEG& 18.8 & 23.32 & 39.10 &  dying/recurrent & BCG in A262\\
B2 0222+36 & C     &LEG& 4.5 & 23.93 &  40.71  & young/restarted& jet instability\\
B2 0258+35 & CSS   &LEG& 1.3 & 24.35 & 40.38  & HI rich &frustrated \\
B2 0648+27 & C     &HEG& 1.2 & 24.02 & 41.58  & HI rich& frustrated \\
B2 0708+32B & FR I &? & 10.2 & 24.49  &- &  young & CSO -like\\
B2 0722+30 & FR I  &?& 13.5 & 23.49  &- &  peculiar & Disk galaxy\\
B2 1037+30 & CSO   &LEG& 5.0 & 25.36  & 40.08  & young & BCG in A923\\
B2 1101+38 & BL-Lac&?  &18 & 24.39  & 40.54  & orientation & MKN421\\
B2 1217+29 & C     &LEG & 0.01 & 21.75  & 38.83  & HI rich& N4278 low power/frustrated\\
3C 272.1   & FR I  &LEG& 14.6 & 23.59 & 38.20  & young &\\
B2 1254+27 & FR I  &LEG& 16.7 & 23.45  & 39.01 & young & BCG group\\
B2 1257+28 & FRI   &LEG&9.6 & 23.81 & 40.86  & frustrated & N4874, BCG in Coma\\
B2 1322+36B & FR I &LEG& 18.7 & 24.07  & 39.42  & young &\\
B2 1346+26 & FR I  &LEG& 13.2 & 25.47  & 40.58  & frustrated/jet instability & BCG A1795\\
3C 305     & FR I  &HEG& 11.2 & 25.44  & 41.05  & frustrated &\\
B2 1557+26 & C     &LEG&3.1 & 24.05  & 39.29  & low power/frustrated &\\
B2 1855+37 & C     &LEG&  10.6 & 24.75  && dying/restarted&\\
\hline                                                           
\hline                                                           
B2 0331+39 & FRI  &In& 70 & 24.19 &-& restarted &\\
B2 0844+31 & FRII &LEG& 383.2 & 25.50  & 40.35 & restarted& IC 2402\\
B2 1003+35 & FRII &LEG& 4475 & 26.12  & 41.07  &restarted & 3C 286\\
B2 1144+35 & FRI  &HEG& 839 & 24.41  & 41.14  &high/low phase&\\
B2 1512+30 & FRI  &LEG& 38 & 24.71  & 40.02  & dying &\\
B2 1626+39 & FRI  &LEG& 61.9 & 25.56  & 39.57  &&3C 338, BCG in A2199\\
\hline
\hline
\end{tabular}
\end{center}
{\scriptsize Col.1: names of the sources are listed; Col.2: radio morphologies; Col.3: optical identification, where ? indicates source for which we could give a classification;\\ Col. 4: Linear size; Col. 5 : Logarithm of total radio power at 408 MHz in units of W/Hz; Col. 6: Log$L_{[OIII]}$ is the
  logarithm of [OIII]$\lambda$ 5007 \AA\ line luminosity in unit of 10$^{-15}$ erg s$^{-1}$; in Col.7, it is reported the classification based on our radio and optical analysis; Col. 8 gives some notes on sources properties.} 
\end{table*}

\item{\bf sources in HI-rich galaxies.\\} 
Results of Emonts et al. 2006a, 2006b revealed that FRI type lie in a 
particular region of the HI mass disk/radio power diagram. However, there are 
sources that differ from FRI type having a large value of HI mass. Studying 
in details these latter objects, they found that all these are compact sources,
even if not all compact sources have large HI mass.  
Some of our C BCS sources (B2 0648+27, B2 0258+35 and NGC 4278, Tab. 6) show a large 
amounts of extended HI disk. For these HI rich compact objects Emonts et al. 
2006a and 2006b suggest that they do not grow into extended sources because 
they are frustrated by the ISM in the central region of the galaxy, or 
because the fuelling stops before the sources can expand. If these H I-rich 
radio galaxies formed through a major merger of gas-rich galaxies, part of 
the gas is expelled in large-scale tidal structures, while another part is 
transported into the central kpc-scale region (e.g. Barnes 2002). The latter 
could be responsible for frustrating the radio jets if they are not too 
powerful. Alternatively, while the geometry and the conditions of the 
encounters appear to be able to form the observed large-scale HI structures, 
they might not be efficient in channeling substantial amounts of gas to the 
very inner pc-scale region. This might prevent stable fuelling of the AGN and 
hence large-scale radio structures do not develop. This hypothesis seems 
reasonable looking through our C BCSs: evidences that NGC 4278 and B2 0648+27 cannot grow
as they are frustrated by the local ISM are present (Giroletti et al. 2005b); while B2 0258+35 displays variable 
levels of activity, suggestive of inefficient fuelling, to expand beyond the kpc scale (Giroletti et al. 2005b).

We found also a slight correlation between the amount of HI mass and the 
central radio morphology: objects with high HI mass (e.g. B2 0258+35) show 
more diffuse radio emission in the central region. 

Emonts et al. 2006a, 2006b discuss that HI-rich low power 
compact sources have different formation history from FRIs objects 
being likely the products of major merger as the detected large amounts of HI 
demonstrate. In this scenario, it will be interesting to note that in B2 0648+27 and B2 1217+29 the presence of a major merger is clearly
confirmed (Emonts et al. 2006a, 2006b).

\end{itemize}

\section{CONCLUSION.} \label{conclusion}
 Radio galaxies are classified as FRI, FRII and Compact sources according to 
their powers and morphologies. Compact objects show emission properties which 
are not yet well 
understood. 
To investigate this peculiar class of sources, we selected from the Bologna 
Complete Sample (BCS, Giovannini et al. 2001, 2005; Liuzzo et al. 2009b) all objects 
with a linear size smaller than 20 kpc, forming the C BCS. Part of these targets were 
previously analyzed by us in the radio band (Giroletti et al. 2005b). Here, 
we complete the radio analysis of the C BCS sample presenting new high resolution VLA observations 
for the remaining sources. Moreover, we discuss for the first time all optical available data for C BCSs.

From the comparison between C BCSs and other source samples/extended radiogalaxies, we derive that:

\begin{itemize}

\item diagnostic diagrams reveal that with a few exceptions,
C BCSs show optical LEG properties as 3CR FR I radio galaxies and CoRGs of Baldi $\&$ Capetti 2010.
 
\item The optical [OIII] - radio correlations (total and nuclear radio power)
suggest a common linear correlation for CoRGs and C BCS 
sources, different from the known linear correlation of HEG and LEG 3CR radio
galaxies, suggesting that C BCSs could be the powerful tail of CoRGs.

A possible continuity with powerful CSS it is not yet clear.

\item From our sub-arcsec radio data, the compactness of C BCSs is
mostly due to a low source age and/or restarted activity in a gas rich
environment (e.g. BCG galaxies and HI-rich galaxies).
Projection effects hold in a very few cases in agreement with unified models 
predictions.

\end{itemize}

\begin{acknowledgements}
 This work was supported by contributions of European Union, Valle D'Aosta Region and the Italian Minister for Work and Welfare. 
This research has made use of the NASA/IPAC Extragalactic Data Base (NED), which is operated by the JPL, California Institute of Technology, under contract with the National Aeronautics and Space Administration.

\end{acknowledgements}

\end{document}